%% file: main.tex
\documentclass[sigconf]{acmart}

\author{Yao Tian}
\affiliation{%
  \institution{The Hong Kong University of \\Science and Technology}
  \city{Hong Kong}
  \country{China}
}
\authornote{Both authors contributed equally to this research.}
\email{ytianbc@cse.ust.hk}
\orcid{0000-0001-6876-5059}
\author{Zhoujin Tian}
\authornotemark[1]
\affiliation{%
  \institution{The Hong Kong University of \\Science and Technology}
  \city{Hong Kong}
  \country{China}
}
\email{ztianaf@cse.ust.hk}
\orcid{0009-0001-7673-7311}

\author{Xi Zhao}
\affiliation{%
  \institution{The Hong Kong University of \\Science and Technology}
  \city{Hong Kong}
  \country{China}
}
\email{xzhaoca@cse.ust.hk}
\orcid{0000-0002-3240-8225}

\author{Ruiyuan Zhang}
\affiliation{%
  \institution{Hong Kong Generative AI Research \\ \& Development Center}
  \city{Hong Kong}
  \country{China}
}
\email{zry@hkgai.org}
\orcid{0000-0003-2022-7387}

\author{Xiaofang Zhou}
\affiliation{%
  \institution{The Hong Kong University of \\ Science and Technology}
  \city{Hong Kong}
  \country{China}
}
\email{zxf@cse.ust.hk}
\orcid{0000-0001-6343-1455}

\AtBeginDocument{%
  }

\setcopyright{none}
\usepackage{geometry}%
\usepackage{graphicx}
\usepackage{subcaption}
\usepackage{caption}
\usepackage{enumerate}
\usepackage{enumitem}
\usepackage{multirow}
\usepackage[ruled,vlined]{algorithm2e}
\usepackage{amsmath, bm}
\newtheorem*{problemstatement}{\noindent\textbf{Problem Statement}}

\newtheorem{definition}{Definition}
\newtheorem{example}{Example}
\newtheorem{remark}{Remark}
\newtheorem{discussion*}{Discussion}

\newcommand{\ie}{\emph{i.e.,}\xspace}
\newcommand{\aka}{\emph{a.k.a.}\xspace}
\newcommand{\eg}{\emph{e.g.,}\xspace}

\usepackage{makecell}

\usepackage{algpseudocode}

\definecolor{myblue}{RGB}{30,90,255}
\def\rone{}
\def\rtwo{} 
\definecolor{mygreen}{RGB}{58,191,153}
\def\rthree{} 
\def\rcommon{}
\usepackage[normalem]{ulem} 
\begin{document}

\title{GEM: A Native Graph-based Index for Multi-Vector Retrieval}



\begin{abstract}

In multi-vector retrieval, both queries and data are represented as sets of high-dimensional vectors, enabling finer-grained semantic matching and improving retrieval quality over single-vector approaches. However, its practical adoption is held back by the lack of effective indexing algorithms. Existing work, attempting to reuse standard single-vector indexes, often fails to preserve multi-vector semantics or remains slow. In this work, we present GEM, a native indexing framework for multi-vector representations. The core idea is to construct a proximity graph directly over vector sets, preserving their fine-grained semantics while enabling efficient navigation. First, GEM designs a set-level clustering scheme. It associates each vector set with only its most informative clusters, effectively reducing redundancy without hurting semantic coverage. Then, it builds local proximity graphs within clusters and bridges them into a globally navigable structure.  To handle the non-metric nature of multi-vector similarity, GEM decouples the graph construction metric from the final relevance score and injects semantic shortcuts to guide efficient navigation toward relevant regions.  At query time, GEM launches beam search from multiple entry points and prunes paths early using cluster cues. To further enhance efficiency, a quantized distance estimation technique is used for both indexing and search. Across in-domain, out-of-domain, and multi-modal benchmarks, GEM achieves up to 16× speedup over state-of-the-art methods while matching or improving accuracy.

\end{abstract}





\maketitle


\input{sec-introduction}
\input{sec-related-work}

\input{sec-preliminary}
\input{sec-method}

\input{sec-experiment}
\input{sec-conclusion}
\begin{acks}	
	The research work described in this paper was supported by Hong Kong Research Grants Council (grant \#16210625, T43-513/23-N), HKUST-WeBank Joint Laboratory (grant \#WEB24EG01), HKUST-MetaX Joint Laboratory for Advanced AI Computing (grant \#META\\X24EG01). It was partially conducted in JC STEM Lab of Data Science Foundations funded by The Hong Kong Jockey Club Charities Trust.
\end{acks}

\bibliographystyle{ACM-Reference-Format}
\bibliography{sample}


\end{document}

%% file: sec-introduction.tex
\section{Introduction} \label{sec:introduction}

\begin{figure}[t]
  \centering
  \includegraphics[width=0.49\textwidth]{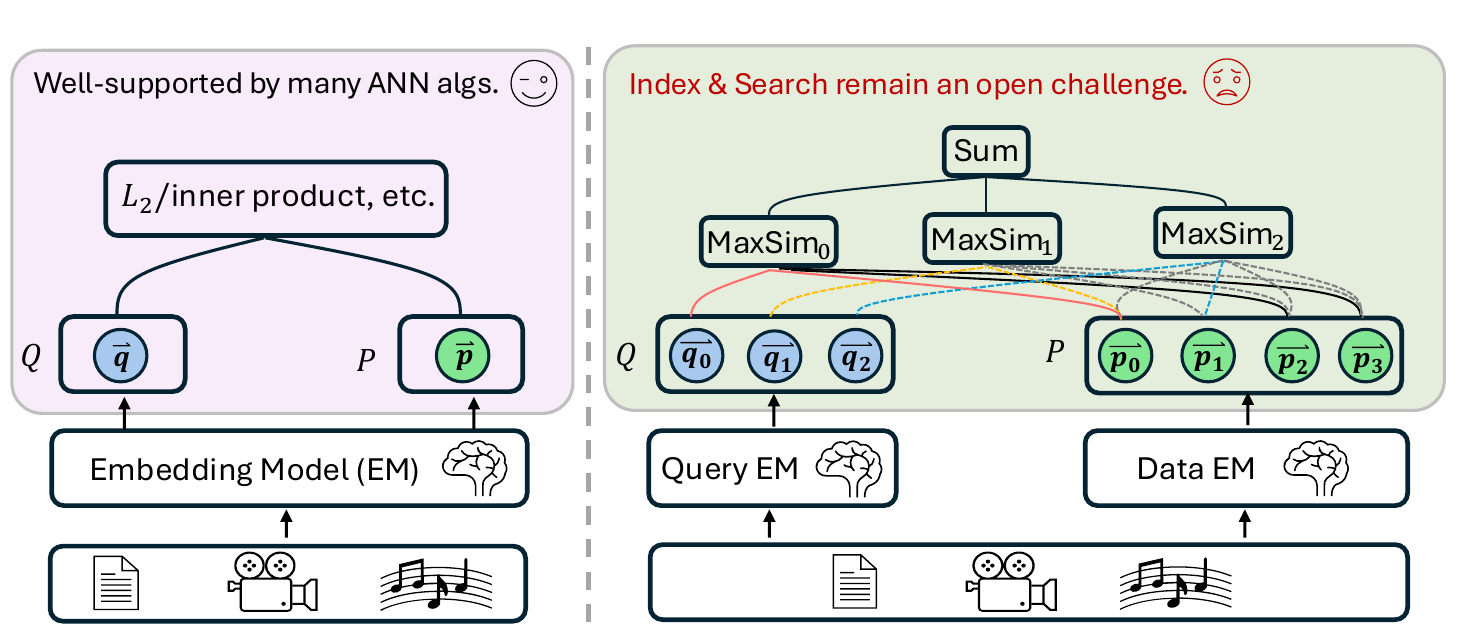}
  \caption{\rthree{Single-Vector v.s. Multi-Vector Retrieval}}
  \label{fig:intro}
\end{figure}

The advent of Large Language Models (LLMs) has profoundly reshaped the information retrieval landscape in recent years. 
\rthree{By embedding diverse data—text, images, videos, and beyond—into high-dimensional vector spaces, LLMs have made vector representations a standard data format, as shown in Figure \ref{fig:intro}(left)}. This shifts the retrieval paradigm from lexical matching to semantic similarity. At its core lies approximate nearest neighbor (ANN) search, which now underpins many modern applications, such as web search \cite{DBLP:conf/emnlp/KarpukhinOMLWEC20, DBLP:conf/kdd/TianL0WSHZHWDXZ23}, recommendation systems \cite{DBLP:conf/kdd/AoualiBBHIHRSVV22,DBLP:conf/www/JiangCH020}, and retrieval-augmented generation (RAG) systems  \cite{DBLP:conf/nips/LewisPPPKGKLYR020, DBLP:conf/icml/0008LSH23,DBLP:journals/corr/abs-2312-10997,DBLP:journals/corr/abs-2305-15334}. However, compressing rich semantics into a single embedding vector—like summarizing an entire passage with a single word—inevitably leads to the loss of fine-grained details. This loss often results in irrelevant retrievals and propagates hallucinations in downstream LLMs, especially in long-context inference scenarios \cite{DBLP:journals/corr/abs-2406-04744}.
To overcome this limitation, modern information systems are increasingly representing complex objects as sets of vectors  \cite{DBLP:conf/sigir/KhattabZ20, DBLP:conf/naacl/SanthanamKSPZ22, DBLP:conf/cikm/HofstatterKASH22, DBLP:conf/nips/LeeDDLNCZ23, DBLP:conf/nips/LinCMCB23, DBLP:conf/iclr/YaoHHLNXLLJX22, DBLP:journals/pacmmod/MollFMGC23}.
ColBERT \cite{DBLP:conf/sigir/KhattabZ20} is a prominent example, which produces multiple embeddings per query or document by generating one embedding per token. The relevance score is then evaluated using the MaxSim operator 
(see Definition \ref{def:chamfer}), which aggregates the maximum similarity of each query embedding to any document embedding. 
As illustrated in Figure \ref{fig:intro} (right), both query and document are represented as sets of vectors, $Q  = \{q_0, q_1, q_2\}$ and $P  = \{p_0, p_1, p_2, p_3\}$, where $q_i, p_j \in \mathbb{R}^d$. Then, each query vector $q_i$ interacts with all vectors in the document $P$: for example, $q_0$ achieves the highest similarity $0.94$ with $p_0$, $q_1$ best matches $p_0$ with score $0.7$, and $q_2$ aligns most with $p_4$ at  $0.22$. The final relevance score is computed by summing or averaging these individual scores, \eg $0.94+0.7+0.22 = 1.86$. This new paradigm has demonstrated significant improvements in retrieval quality \cite{DBLP:conf/ictir/WangMTO21, DBLP:journals/tacl/KhattabPZ21,DBLP:conf/acl/ThaiCKI22, DBLP:conf/nips/Thakur0RSG21}, interpretability \cite{DBLP:conf/ecir/FormalPC21, DBLP:conf/sigir/WangMTO23}, and generalization ability 
\cite{DBLP:conf/ecir/LupartFC23, DBLP:conf/cikm/ZhanXMLGZM22} across various IR benchmarks, emerging as a core functionality in current industrial vector database systems, such as Weaviet \cite{weaviate-multivector}, Vespa \cite{vespa-multivector} and Pinecone \cite{pinecone-multivector}. 

However, these gains come at steep costs in storage, computation, and the complexity of indexing and retrieval. First, representing each item with multiple vectors inflates storage requirements by several orders of magnitude. Second, pairwise distance calculations across all vectors in the set-to-set comparison incur higher costs. These issues can be alleviated by low-bit quantization techniques \cite{DBLP:journals/pacmmod/GaoL24, DBLP:journals/pacmmod/GaoGXYLW25} or by strategically retaining only the most informative vectors \cite{DBLP:journals/corr/abs-2505-19189, DBLP:conf/cikm/HofstatterKASH22}; however, accurate and efficient indexing and search algorithms for multi-vector representation remain an open challenge.
To address this, one straightforward idea \cite{DBLP:conf/sigir/KhattabZ20, DBLP:conf/naacl/SanthanamKSPZ22, PLAID, EMVB, IGP} is to pool all vectors from all vector sets into a single ANN index, \eg IVFPQ \cite{DBLP:journals/pami/JegouDS11} or diskANN \cite{jayaram2019diskann}. 
At query time, each query vector retrieves its top-$k$ NNs, generating the initial candidate pool for refinement.  
However, this approach leads to two issues: 
(1) vector-level similarity does not always imply set-level relevance, resulting in a large number of irrelevant candidates for refinement. For example, a document about "rocky beaches in Portugal" may be retrieved for the query "family-friendly beaches in Europe" due to the shared term "beaches" despite the clear mismatched intent; 
(2) indexing all individual vectors bloats the index. For example, $10k$ documents with $100$ vectors each lead to a million-level ANN task. 
Another proposal \cite{MUVERA}, from Google, attempted to map the vector set into a single vector using distance-preserving mappings, enabling the use of existing ANN indexes. 
However, a theoretically sound distance-preserving mapping requires prohibitively high dimensionality of the projection space that even well-optimized ANN indexes struggle to handle efficiently.  
Moreover, this approach contradicts the motivation for using multi-vector representations and begs the question of why not directly train a single-vector embedding model from the outset?
DESSERT \cite{DESSERT} is the first solution designed natively for multi-vector retrieval.  It compresses the vector set into a sketch via locality-sensitive hashing (LSH) \cite{DBLP:conf/stoc/Charikar02, DBLP:conf/compgeom/DatarIIM04}, and retrieves top-$k$ candidates by comparing hash codes.  However, without effective sketch-level pruning, the speedup is limited, and storage pressure increases due to the large number of hash values.



Motivated by the aforementioned limitations, we propose GEM, a native \underline{\textbf{G}}raph-based ind\underline{\textbf{E}}x framework for \underline{\textbf{M}}ulti-vector retrieval. The key idea of GEM is set-level indexing: constructing a proximity graph where each vertex represents a vector set, in the hope of preserving the semantic richness of multi-vector representations while leveraging the strong pruning capability of graph-based indexes.
Specifically,  GEM incorporates a series of co-designed architectural and algorithmic optimizations.
First, GEM designs a clustering scheme for vector sets. It evaluates the informativeness of the vectors in each set using a TF-IDF–style score and adaptively assigns the set only to the clusters of $r$ highest-scoring vectors. 
This design preserves semantic coverage while avoiding excessive redundancy in the subsequent graph construction. 
Next, GEM proposes a dual-graph structure consisting of intra-cluster graphs and a global graph. The intra-cluster graphs capture compact local structures, while sets shared across clusters naturally form bridges, yielding a globally connected graph that enables efficient cross-region traversal and pruning. 
To reconcile the non-metric nature of the MaxSim operator with the metric assumption underlying graph-based search, GEM decouples the similarity measures used for graph construction and search. 
Specifically, the graph is built using Earth Mover's Distance (EMD) \cite{emd}. 
As EMD is a well-established metric that satisfies the triangle inequality and upper-bounds the MaxSim, the EMD-based graph could provide a stable and reliable basis for navigating toward high-quality results. 
In addition, GEM introduces semantic shortcuts to bridge the gap between the two similarity measures. These shortcuts directly connect vertices that are semantically similar but distant in the graph, accelerating search and helping escape local optima. 
Furthermore, GEM employs quantized versions of EMD and MaxSim. This replaces expensive set-to-set distance calculations with efficient lookups over a centroid codebook, significantly enhancing both indexing and querying efficiency. 
Finally, a cluster-guided multi-entry search algorithm is developed on top of the proposed index. The search begins from multiple entry points within multiple promising clusters and expands over the graph in parallel. 
Meanwhile, a cluster-guided pruning mechanism prevents the search from drifting into irrelevant regions, effectively balancing speed and accuracy.

In summary, we make the following contributions:
\begin{itemize}[leftmargin = 9pt]
    \item  We propose GEM, the first native graph index for multi-vector retrieval.
    It successfully unifies the semantic richness of multi-vector representations with the pruning efficiency of graph-based indexing, delivering both high accuracy and efficiency.
    \item GEM introduces a series of architectural and algorithmic innovations tailored to the unique challenges of multi-vector search. The architecture adopts a dual graph design with metric decoupling and semantic shortcuts, ensuring a navigable graph topology for accurate search. These are integrated with a novel semantic clustering algorithm, a cluster-aware multi-entry search algorithm, and quantized distance estimation, achieving full end-to-end efficiency across large-scale vector-set collections.
    \item  Extensive experiments on in-domain, out-of-domain, and multi-modal benchmarks show that GEM significantly outperforms the state-of-the-art methods in both retrieval quality and latency, with a comparable or even smaller index size and indexing time. 
\end{itemize}


The rest of the paper is organized as follows. Section \ref{sec: preliminary} defines the problem and introduces a straightforward graph-based solution. Our GEM framework is detailed in Section \ref{sec: method}, followed by experimental results in Section \ref{sec:exp}. Section \ref{sec: relatedwork} reviews related work, and we conclude in Section \ref{sec:conclusion}. 

%% file: sec-related-work.tex
\section{Related Work}\label{sec: relatedwork}

\subsection{Single-Vector ANN Search}
The ANN search problem has been extensively studied in the database community for decades \cite{DBLP:journals/debu/TianYZ0Z023}. However, nearly all well-known methods make the single-vector representation assumption, \ie both queries and data are encoded as single high-dimensional vectors in the same Euclidean space.  Prominent approaches include LSH-based methods \cite{DBLP:conf/sigmod/LeiHKT20, DBLP:conf/icde/TianZZ22, DBLP:journals/tkde/TianZZ24, DBLP:journals/pvldb/HuangFZFN15, DBLP:journals/pvldb/LuWWK20, DBLP:journals/pvldb/ZhengZWHLJ20, DBLP:journals/vldb/LiuW00Q021}, quantization-based methods \cite{DBLP:journals/pami/JegouDS11, DBLP:journals/pami/GeHK014, DBLP:journals/pacmmod/GaoL24,  DBLP:journals/pacmmod/GaoGXYLW25}, and graph-based methods \cite{DBLP:journals/pami/MalkovY20, DBLP:journals/pvldb/FuXWC19, DBLP:journals/pr/MunozGDT19,jayaram2019diskann,DBLP:journals/pvldb/ZhaoTHZZ23,DBLP:journals/pacmmod/WangXYWPKGXGX24}. LSH-based methods are well-known for the strong theoretical guarantees on recall. Quantization-based methods significantly reduce memory usage and accelerate distance calculations. Graph-based methods, which construct a proximity graph among vectors and find neighbors via greedy traversal, have become the de facto state-of-the-art, demonstrating superior performance in both secch accuracy and efficiency.

\subsection{Multi-Vector ANN Search}
Multi-vector search is a relatively new paradigm that has seen rapid adoption across various domains due to its high effectiveness \cite{DBLP:conf/sigir/KhattabZ20, DBLP:conf/nips/KhattabPZ21, DBLP:conf/iclr/ParanjapeKPZM22, DBLP:conf/naacl/LiFSILS22}. \rtwo{Unlike single-vector approaches, which alleviate information loss by increasing dimensionality but exacerbate the curse of dimensionality, multi-vector models capture complex semantics from a different angle: they represent an object with multiple simpler (\eg not very high-dimensional) vectors.}
It is first introduced for document retrieval by ColBERT \cite{DBLP:conf/sigir/KhattabZ20}, which represents queries and documents as sets of token-level embeddings and employs a \textit{late interaction} architecture to achieve strong performance. Subsequent work further improved the effectiveness of this paradigm, for instance, by learning different weights for vectors of varying importance \cite{DBLP:conf/cikm/HofstatterKASH22}, or expanding it to other domains such as visual-language retrieval tasks \cite{DBLP:conf/acl/LinMCB24, DBLP:conf/iclr/YaoHHLNXLLJX22}. 
However, early research relies on simple inverted lists to index all individual embeddings, which hinders large-scale deployment.
To improve scalability, ColBERTv2 \cite{DBLP:conf/naacl/SanthanamKSPZ22} introduces a centroid-based compression scheme, where each token embedding is stored using the ID of its closest centroid and a low-bit residual (1–2 bits per component), significantly reducing storage overhead. However, the search process remains costly. PLAID \cite{PLAID} builds on this by further exploiting the centroids to aggressively prune irrelevant documents via a \textit{centroid interaction }mechanism, substantially accelerating retrieval. EMVB \cite{EMVB} enhanced PLAID with system-level optimizations, including bit-vector filtering and SIMD-accelerated scoring. More recently, IGP \cite{IGP} proposes to  replace the inverted list with a proximity graph index over centroid vectors and develops an incremental next-similar retrieval technique to efficiently filter low-relevance candidates.
Alternative indexing strategies have also been explored.  
MUVERA \cite{MUVERA} introduces fixed-dimensional encodings (FDEs) that approximate multi-vector similarity, enabling the use of optimized single-vector ANN indexes. While efficient, this method inevitably loses token-level alignment, risking reduced recall. DESSERT \cite{DESSERT} proposes compressing vector sets into compact sketches, but its speedup is often limited by a lack of effective set-level pruning.
Orthogonal to the research on indexing and search algorithms, XTR \cite{DBLP:conf/nips/LeeDDLNCZ23} improves retrieval efficiency by simplifying the scoring mechanism during training. CITADEL \cite{DBLP:conf/acl/LiLOGLMY023} reduces computation cost by selectively considering only a subset of tokens. 
POQD \cite{DBLP:journals/corr/abs-2505-19189} enhances accuracy by optimizing query decomposition with the assistance of LLMs. Since these approaches optimize the set-level representation itself,  they can complement the above indexing methods for higher search accuracy and efficiency.

%% file: sec-preliminary.tex
\section{Preliminary}\label{sec: preliminary}
In this section, we formulate the multi-vector retrieval problem and introduce a baseline solution.

\subsection{Problem Definition}
We first clarify the similarity measurement used throughout this paper. Following ColBERT \cite{DBLP:conf/sigir/KhattabZ20} and other state-of-the-art multi-vector methods \cite{DBLP:conf/sigir/KhattabZ20, DBLP:conf/acl/LinMCB24, PLAID, DESSERT, MUVERA, IGP},   we adopt the MaxSim operator, \aka Chamfer similarity, to compute the similarity score between two multi-vector sets. In the following, we choose the terminology Chamfer (CH) due to its historical precedence \cite{DBLP:conf/nips/BakshiIJSW23, MUVERA}.
\begin{definition}[Similarity score]\label{def:chamfer}
Given two sets of vectors $A, B \subseteq \mathbb{R}^d$, the similarity score is defined as the sum of the maximum similarity between each vector in $A$ and all vectors in $B$. Formally, 
\begin{equation}
\label{eq:ch}
    \textsf{CH}(A, B) = \sum_{a\in A} \max_{b\in B} \text{Sim}(a, b),
\end{equation}
where cosine similarity or $L_2$ distance are typically used as the default choices for $\text{Sim}(\cdot, \cdot)$ in prior work.
\end{definition}


\noindent\begin{problemstatement}[multi-vector retrieval problem]
    Given a database $\mathcal{D} = \{P_1, P_2, \ldots, P_N\}$, where each $P_i = \{p_1, p_2, \ldots, $ $ p_{m_i}\}$ is a set of vectors with each $p_j \in \mathbb{R}^d$, and a query $Q = \{q_1, q_2, \ldots, $ $q_{m_q}\}$ with each $q_j \in \mathbb{R}^d$, the multi-vector retrieval problem, \aka the vector set search problem, aims to return the top-$k$ sets in $\mathcal{D}$  that are most similar to $Q$, namely:
\begin{equation}
    \mathcal{R} = k\text{-}\arg\max\{\textsf{CH}(Q,P)|P\in \mathcal{D}\}.
\end{equation} 
\end{problemstatement}
\begin{remark} 
    Multi-vector retrieval has been widely adopted in various domains and systems \cite{DBLP:conf/cvpr/Reddy0YY0MKMDC25, DBLP:conf/sigir/KhattabZ20, DBLP:conf/nips/KhattabPZ21, DBLP:conf/iclr/ParanjapeKPZM22, DBLP:conf/naacl/LiFSILS22}. A prominent example is ColBERT \cite{DBLP:conf/sigir/KhattabZ20} for document retrieval, where queries and documents are encoded into sets of token-level vectors. Throughout this work,  we use the terms "document" and "vector set" interchangeably, as well as "token" and "vector" interchangeably.  However, it is worth noting that multi-vector retrieval itself is a general problem, agnostic to the data's origin. A "vector" could represent an image patch, a chunk of text, or other forms of data.
\end{remark}

\subsection{A Baseline Solution}
\label{sec:naivesolution}
Recall the ANN search problem in the traditional single-vector setting: both data and queries are represented as high-dimensional vectors, and the goal is to find the top-$k$ most similar (\ie closest) data points to a given query based on certain distance functions (\eg $L_2$ distance). Formally, given a dataset $\mathcal{D} = \{o_1, ..., o_N\}$ with each data object $o_i \in \mathbb{R}^d$ and a query $q\in \mathbb{R}^d$, it aims to  return
\begin{equation}
   \mathcal{R} = k\text{-}\arg\min\{\|q, o_i\|_2|o_i\in \mathcal{D}\}.
\end{equation}
Among ANN solutions, graph-based methods have demonstrated the best trade-off between search accuracy and efficiency on real-world datasets. These methods organize the base vectors into an approximate proximity graph (APG), where each vertex corresponds to a data vector, and edges connect vertices that are sufficiently similar. During query time, a greedy (beam) search is performed on the graph: it starts from an arbitrary vertex and iteratively traverses edges toward neighbors that are closer to the query, until convergence to a local optimum.
Inspired by the success of graph-based indexes in the single-vector setting, we pose a question: \textbf{Can we build a set-level APG?} 
In such a graph, each vertex represents a vector set, and an edge connects two sufficiently similar vertices based on the Chamfer distance. 
The graph construction and search processes are the same as those in single-vector methods, so we omit them here for brevity.
Unfortunately, this direct extension does not work due to two critical issues:

\begin{itemize}[leftmargin = 15pt]
\item[(1)] Fragmented neighbors and local optima.
Chamfer distance is not a metric and lacks triangle inequality, which undermines core assumptions behind graph-based search. For example, even if $Q$ is close to $P_1$ and  $P_1$ is close to  $P_2$, $Q$ may still be far from  $P_2$. As a result, the nearest neighbors of $Q$ are scattered across the graph, making them difficult to reach via local hops and slowing down convergence. Without strong connectivity among these neighbors, greedy search paths become oscillatory and unstable, often getting trapped in local optima—terminating early even when better candidates exist several hops away, ultimately degrading recall.
 
\item[(2)] Lack of cheap distance approximation and pruning.
While graph structures provide a powerful pruning framework, the exorbitant cost of computing the Chamfer distance remains a critical bottleneck. It is impractical to perform set-to-set comparisons between the query and all candidate neighbors at each step of the search traversal. A practical solution, therefore, necessitates a cheaper distance approximation, effective pruning strategies, and more informed entry points.

\end{itemize}
We will empirically validate these analyses in Section \ref{sec:exp}, where this naive approach, named MVG, is included as a baseline.

%% file: sec-method.tex
\section{\textsf{GEM}}\label{sec: method}

To address the limitations of existing approaches and the challenges exposed by the naive graph-based solution, we propose GEM — a native \underline{\textbf{G}}raph-based ind\underline{\textbf{E}}xing framework for efficient \underline{\textbf{M}}ulti-vector retrieval.
 The key innovations are as follows:
\begin{itemize}[leftmargin = 15pt]
\item[(1)] We propose a set-level clustering scheme, where each document is assigned to only a small number of semantically relevant clusters, guided by adaptive TF-IDF–style weighting over centroids. This reduces index redundancy while preserving semantic coverage (Section \ref{sec:cluster} and Section \ref{sec: adaptivek}).

\item[(2)] We design a novel dual graph structure, featuring both intra-cluster graphs and a unified global graph naturally bridged by sets shared between clusters. This design guarantees global connectivity while preserving local structure (Section \ref{sec:graphconstruction}).

\item[(3)] We decouple the similarity measures used for graph construction and search: Earth Mover’s Distance (EMD) \cite{emd} is used for graph construction due to its metric properties, while Chamfer distance is retained for search. Since EMD upper bounds Chamfer, vertices that are close under EMD are also close under Chamfer, ensuring reasonable navigation. We further augment the graph with shortcuts, directly linking semantically close pairs under Chamfer, but may require several hops under EMD, which improves search efficiency and helps escape local optima (Section \ref{sec: cluserAPG} and Section \ref{sec: shortcuts}).

\item[(4)] We introduce quantized versions of EMD and Chamfer to alleviate the high cost of set-to-set similarity computation. By leveraging centroid-level codebooks and quantized representations derived from cluster centroids, we improve the efficiency of both the indexing and search phases (Sections \ref{sec:quant} and Section \ref{sec: query}).

\item[(5)] We devise a new cluster-guided, multi-entry search strategy. The search is initialized from multiple semantically relevant clusters, enabling parallel, layer-wise expansion over the global graph. A cluster-guided early pruning mechanism aggressively culls unpromising paths, preventing the search from drifting into irrelevant regions (Section \ref{sec: query}).
\end{itemize}

\begin{figure}[t]
  \centering
  \includegraphics[width=0.46\textwidth]{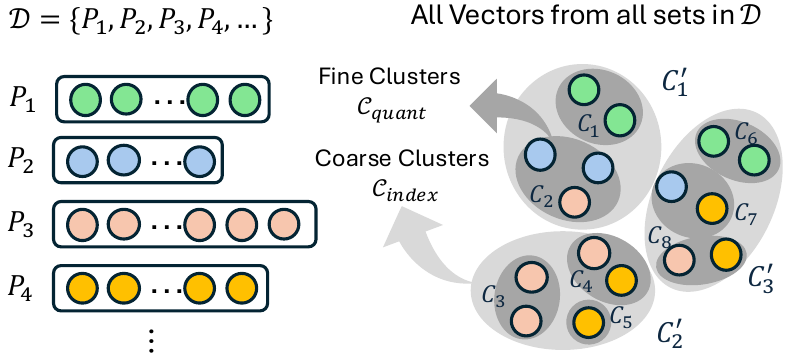}
  \caption{Two-Stage Clustering}
  \label{fig:two-stage}
\end{figure}

\begin{figure}[t]
  \centering
  \includegraphics[width=0.46\textwidth]{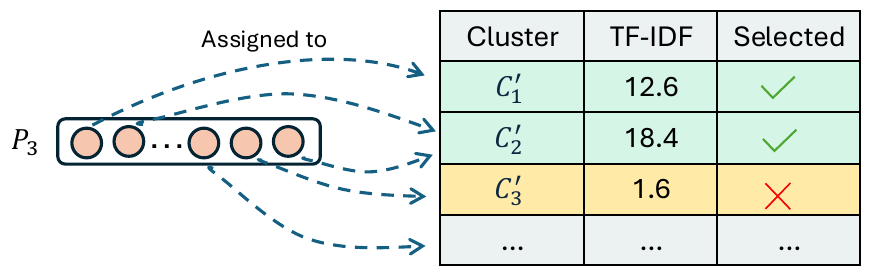}
  \caption{Cluster Assignment with TF-IDF}
  \label{fig:tfidf}
\end{figure}

\subsection{Set-Level Clustering}
\label{sec:cluster}
\subsubsection{Two-Stage Clustering}
\label{sec:twostagecluster} \rone{To facilitate efficient filtering and quantization, we begin with a two-stage clustering step.
In the first stage, we perform k-means clustering on a sample of vectors across all sets in $\mathcal{D}$ to obtain a set of centroids $\mathcal{C}_\text{quant} = \{C_1, C_2, \dots, C_{k_1}\}$.}  These centroids serve as a \textit{vocabulary} for our subsequent quantized distance approximations, \ie qEMD during graph construction to be introduced in Section \ref{sec:quant} and qCH during search in Section \ref{sec:qCH}.
In the second stage, we cluster $\mathcal{C}_\text{quant}$ into a smaller set of coarser centroids $\mathcal{C}_\text{index} = \{C'_1, C'_2, \dots, C'_{k_2}\}$, which defines a cluster space for set-level indexing. 
As illustrated in Figure \ref{fig:two-stage}, all vectors are first grouped into 8 fine-grained clusters, which are then organized into 3 coarse clusters that serve as higher-level semantic bins. Then, we assign each vector $p \in P$ to its nearest cluster centroid in $\mathcal{C}_\text{index}$, denoted as $\text{NN}(p)$.
As a result, a set $P = \{p_1, p_2, \dots, p_{m}\}$ is approximately represented as a set of centroids:
\begin{equation}
    \small
    \mathcal{C}(P) = \{\text{NN}(p_1), \text{NN}(p_2), \dots, \text{NN}(p_{m})\}.
\end{equation}

To obtain a set-level clustering, a naive way is to associate a set with every cluster if one or more of its tokens are assigned to that centroid. For example, $P_3$ is linked to all three clusters $C'_1, C'_2, C'_3$. However, this can cause a vector set to appear in an excessive number of clusters, especially when it contains uninformative tokens, such as "is", "the", and "of". This leads to index redundancy and increased query-time overhead. 
\rtwo{Motivated by the unequal semantic contribution of tokens, we introduce a TF-IDF-guided pruning strategy in the following. }

\subsubsection{TF-IDF Guided Cluster Pruning} 
\label{sec: tfidfpruning} \rtwo{The key idea is that: retain only clusters associated with \textit{important} tokens, while ignoring \textit{uninformative} ones. This way, a set is expected to be assigned to only a few \textit{salient} clusters, which help reduce index redundancy while preserving the semantic richness of multi-vector representations.} To achieve this, we design a TF-IDF–style importance score over the centroid vocabulary $\mathcal{C}_\text{index}$  to reflect how salient the cluster is to the set $P$. Specifically,  we define \textit{"term frequency"} (TF) as:
\begin{equation}
\small
    \text{TF}(C'_j, P) = |\{p \in P : \text{NN}(p) = C'_j, C'_j \in \mathcal{C}_\text{index}\}|,
\end{equation}
which counts how many token vectors in $P$ are associated with a cluster $C_j$. Then, we apply an \textit{"inverse document frequency"} (IDF) to downweight clusters shared across many sets:
\begin{equation}
\small
    \text{IDF}(C'_j) = \log \frac{N}{1 + |\{P' \in \mathcal{D} : \text{TF}(C'_j, P') > 0\}|},
\end{equation}
where $N$ is the number of vector sets in the database. The overall TF-IDF score is:
\begin{equation}
\small
    \text{Score}(C'_j, P) = \text{TF}(C'_j, P) \cdot \text{IDF}(C'_j).
\end{equation}
We then construct a weighted centroid profile for each document:
\begin{equation}
\small
    \mathcal{S}(P) = \left[ (C'_{j_1}, s_{j_1}),\, (C'_{j_2}, s_{j_2}),\, \dots,\, (C'_{j_m}, s_{j_m}) \right],
\end{equation}
where $s_{j_t} = \text{Score}(C'_{j_t}, P)$ indicates how strongly cluster $C'_{j_t}$ represents the semantics of $P$. Based on this profile, we assign $P$ to the top-$r$ clusters with the highest TF-IDF scores:
\begin{equation}
    \small
    \mathcal{C}_{\text{top}}(P) = \text{Top-}r_{C'_j \in C(P)}\, \text{Score}(C'_j, P).
\end{equation}
Figure \ref{fig:tfidf} provides a concrete example of this pruning. For $P_3$, our TF-IDF guided mechanism identifies that its connection to cluster $C'_3$ is weak (score of $1.6$). Therefore, only the two clusters with high semantic relevance (scores $18.4$ and $12.6$) are selected as  its indexing targets. This pruning step ensures that each document is linked only to its most representative clusters, improving both graph construction efficiency and candidate generation efficiency.

\begin{figure}[t]
  \centering
  \includegraphics[width=0.4\textwidth]{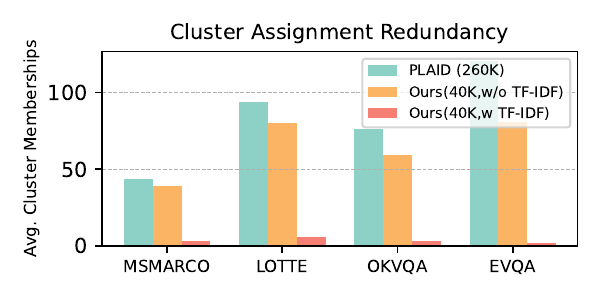}
   \vspace{-3ex}
  \caption{Average \# of Clusters Each Set is Assigned to}
  \vspace{-2ex}
  \label{fig:cluster_membership}
\end{figure}

\begin{remark}
    Clearly, there is no single $r$ value that works best for all: some sets are well represented by a few clusters, while others may require broader cluster coverage. We defer the details of how to determine $r$ adaptively to Section \ref{sec: adaptivek}. 
\end{remark}

\begin{discussion*}
    PLAID adopts a similar clustering idea, but differs from our approach in two aspects. It performs only fine-grained token-level clustering and directly assigns each document to every cluster that any of its tokens maps to. However, we found this strategy less suitable for our graph-based indexing framework. In particular, PLAID's large number of fine-grained clusters results in very few documents per cluster (\eg often only dozens), making it unnecessary and inefficient to build a graph index per cluster.  Moreover, the high redundancy in cluster memberships results in costly graph construction and many redundant edges during query processing. Therefore, to better support our graph-based design, we adopt the fine-grained clustering in the first stage for accurate distance approximation, while second-stage coarser clusters, combined with TF-IDF pruning, significantly reduce per-document cluster memberships, striking a better balance between granularity and efficiency.
\end{discussion*}

Figure~\ref{fig:cluster_membership} compares the average number of clusters each document is assigned to under different strategies across four datasets. The \texttt{PLAID} bars correspond to PLAID’s default configuration with $260K$ fine-grained clusters, resulting in high redundancy. The \texttt{Ours (40K, w/o TF-IDF)} bars represent a coarser 40K-cluster setup without applying TF-IDF guided cluster filtering, while the \texttt{Ours (40K, w/ TF-IDF)} bars show the results after pruning.
As shown, our approach dramatically reduces the number of cluster memberships, \eg from 43.8 to 2.9 on MSMARCO (a 93.4\% reduction), from 94 to 5.8 on LOTTE (93.8\%), from 76 to 3 on OKVQA (96\%), and from 120.8 to 1.7 on EVQA (98.6\%), alleviating index redundancy while preserving semantic coverage.

\subsection{Metric Decoupling and Quantization}
\subsubsection{Metric Decoupling}
\label{sec: decoupling}

\begin{figure*}[t]
\vspace{-1ex}
  \centering
  \includegraphics[width=0.8\textwidth]{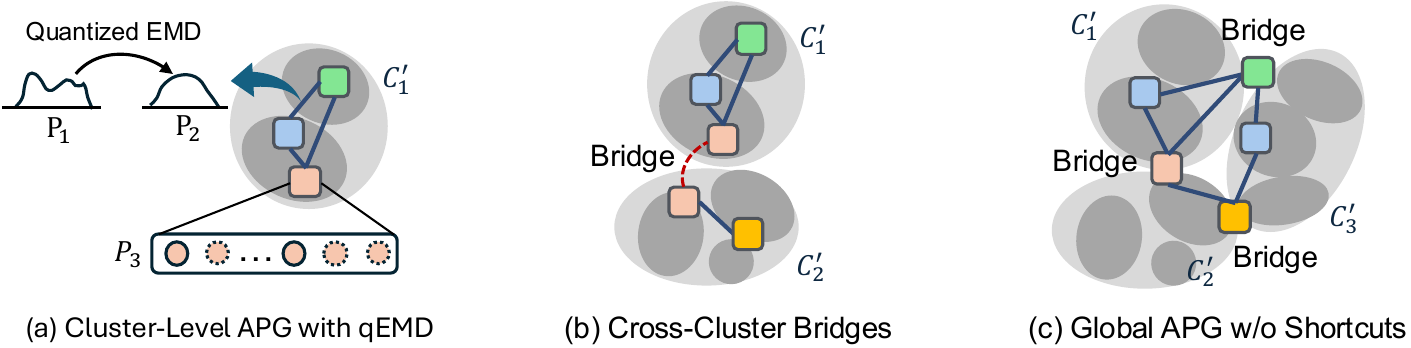}
  \vspace{-1ex}
  \caption{Overview of Global Graph Construction}
  \vspace{-1ex}
  \label{fig:globalgraph}
\end{figure*}
\rtwo{As discussed in Section \ref{sec:naivesolution}, multi-vector similarity lacks the metric properties required to serve as a reliable foundation for building a stable and navigable APG.}
To address this issue, we propose using EMD \cite{emd} as the edge weight metric, while performing search based on the original Chamfer distance. EMD is a well-established distance function that satisfies the triangle inequality, ensuring a well-structured graph topology with guaranteed monotonicity. A key insight supporting this metric decoupling is the bounding relationship between EMD and Chamfer distance. It can be proven that for any two vector sets \( Q \) and \( P \):
\begin{equation}
\small
    \text{CH}(Q, P) \leq \text{EMD}(Q,P).
\end{equation}
This inequality implies that if two sets are close under EMD, they are at least as close--if not closer--under Chamfer distance. Consider a search path that is currently at vertex $P_1$, then the distance from $P_1$ to its neighbor $P_2$ has a predictable upper bound: $CH(Q, P_2) \leq EMD(Q, P_2) \leq EMD(Q, P_1) + EMD(P_1, P_2)$. Consequently, navigating toward the nearest neighbor under EMD is a reliable heuristic for finding a reasonable candidate in the Chamfer space. This mitigates oscillatory search behavior and the local optima trap. EMD is formally defined as follows: given two vector sets \( P_1 = \{p_1, \dots, p_{m_1}\} \) and \( P_2 = \{p'_1, \dots, p'_{m_2}\} \), and a transport plan \( T = \{t_{ij}\} \), then
\begin{equation}
\small
    \text{EMD}(P_1, P_2) = \min_T \sum_{i=1}^{m_1} \sum_{j=1}^{m_2} t_{ij} \cdot d_X(p_i, p'_j)
\end{equation}
subject to the following constraints:
\begin{align}
\small
    \sum_{j=1}^{m_2} t_{ij} = \frac{1}{m_1}, \quad
    \sum_{i=1}^{m_1} t_{ij} = \frac{1}{m_2}, \quad
    t_{ij} \geq 0
\end{align}
where $d_X(p_i, p'_j)  = 1 - \langle  p_i, p_j' \rangle$ when cosine similarity is used, and $d_X(p_i, p'_j)  = \| p_i, p_j' \|_2$ when the $L_2$ distance is adopted.

\subsubsection{Quantization}
\label{sec:quant}
While EMD offers desirable properties for graph construction, its exact computation is prohibitively expensive \cite{DBLP:conf/soda/AndoniIK08, DBLP:conf/stoc/Charikar02, DBLP:conf/cikm/LvCL04,DBLP:conf/eurosys/LvJWCL06,DBLP:conf/nips/GraumanD06}. To make EMD practical at scale, we introduce an efficient quantized approximation. \rone{The key idea is to replace expensive token-level distance computations in EMD with efficient centroid-level approximations by leveraging the shared centroid vocabulary  $ \mathcal{C}_\text{quant}$ derived from k-means clustering (Section \ref{sec:twostagecluster}). Specifically, each token vector is first mapped to its nearest centroid in $\mathcal{C}_\text{quant}$, and the distance between raw vectors is then approximated by the distance between their corresponding assigned centroids, formalized as:
\begin{equation}
\small
     d_X(p_i, p'_j) \approx d_X(\text{NN}(p_i), \text{NN}(p'_j)),
\end{equation}
where \( \text{NN}(p_i) \) and \( \text{NN}(p'_j) \) $ \in \mathcal{C}_\text{quant}$. This quantization brings two key computational benefits. First, all centroid-to-centroid distances can be precomputed and stored in a \( k_1 \times k_1 \) codebook, enabling constant-time distance lookups instead of costly high-dimensional distance calculations. Second, the reduced centroid space generates significantly sparser transport plans, which further accelerate EMD computation.} Formally, the quantized EMD (qEMD) is as follows:
\begin{equation}
\small
    \text{qEMD}(P_1, P_2) = \min_T \sum_{i=1}^{n_1} \sum_{j=1}^{n_2} t_{ij} \cdot d_X(\text{NN}(p_i), \text{NN}(p'_j)).
\end{equation}
\begin{algorithm}[t]
    \caption{Index Construction Pipeline}
    \label{algo:overall-index}
    \LinesNumbered
    \small
    \KwIn{Database $\mathcal{D}$, two-stage cluster counts $k_1, k_2$, degree limit $M$,  parameter $f$, training pairs $\mathcal{T}$,}
    \KwOut{GEM proximity graph $G$}
    Initialize $G$ with $V = \mathcal{D}$ and $E = \emptyset$\;
    $\mathcal{C}_{\text{quant}}, \mathcal{C}_{\text{index}}, \mathcal{C}_{\text{top}}\; \gets$ \Call{ClusterAndAssign} {$\mathcal{D}, k_1, k_2$} \label{ln:clusterandassign}\; 
     \ForEach{$C_i \in \mathcal{C}_{\text{index}}$}{
        $G \gets$ \Call{LocalGraphandBridges}{$C_i, \mathcal{C}_{\text{quant}}, \mathcal{C}_{\text{top}}, f, M$}; \hfill ~\Comment{\textit{Alg. \ref{algo:cluster-graph}}}  \label{ln:callclusterindex}\\
     }
    $G \gets$ \Call{InjectShortcuts}{$G, \mathcal{T}, M$};\Comment{\textit{Alg. \ref{algo:inject-shortcuts}}} \\
    \Return{$G$}
\end{algorithm}

\subsection{Graph Construction}
\label{sec:graphconstruction}
Building upon the set clustering and distance function introduced earlier, this section details the construction of GEM's dual-graph structure. Algorithm~\ref{algo:overall-index} outlines the indexing pipeline.
\subsubsection{Cluster-Level APG with qEMD}
\label{sec: cluserAPG} For each cluster $ C_i$, we invoke  \textsc{LocalGraphandBridges} (Line \ref{ln:callclusterindex} in Algorithm \ref{algo:overall-index}) to build a cluster-level graph while handling cross-cluster bridges.  Specifically, for each vector set $P \in C_i$, we find its top-$f$ ANNs by performing the search algorithm (Algorithm \ref{algo:query}) over the current cluster graph, starting from an arbitrary entry point in $C_i$ and using qEMD as the distance metric (Line \ref{line: findANNs} in Algorithm \ref{algo:cluster-graph}). If $P$ is new to the current graph $G$ (Line \ref{line: new}), we directly connect it to its neighbors (Line \ref{line: connectedges}) and ensure that each neighbor $P'$ does not exceed the degree limit $M$ by removing the least similar edges if necessary (Line \ref{line: remove}). Otherwise,  \( P \)  has already been inserted in previous clusters \( \{C_i \mid i < j,\, C_i \in \mathcal{C}_{\text{top}}(P)\} \), which means $P$ can serve as a cross-cluster bridge. In this case, we invoke \textsc{UpdateBridges}  (Algotihrm \ref{algo:bridge-update}) to merge the old and new neighbor candidates (Line \ref{line: refertoupdatebridge}), as detailed in the next section.

\begin{algorithm}[t]
    \caption{Build Cluster Graph with Cross Bridges}
    \label{algo:cluster-graph}
    \LinesNumbered
    \small
    \KwIn{Cluster $C_i$, current global graph $G = (V, E)$, $\mathcal{C}_\text{top}(\cdot)$, $f$, $M$}
    \KwOut{Updated global graph $G$}
    
    \ForEach{$P \in C_i$}{
        $C \gets$ $f$ ANNs of $P$ in the current graph of $C_i$ by qEMD; \Comment{\textit{Alg. \ref{algo:query}}}\label{line: findANNs}\\
        
        \uIf{$E[P] = \emptyset$ \label{line: new}}{
            \ForEach{$P' \in C$}{
                Add a bidirectional edge $(P, P')$ to $E$\; \label{line: connectedges}
                \If{degree of $P' > M$ \label{line: exceedM}}{
                    Remove the least similar edge from $E[P']$\; \label{line: remove}
                }
            }
        }
        \Else{
            $G \gets$ \Call{UpdateBridges}{$P, C, G, \mathcal{C}_\text{top}(P), M$}; \Comment{\textit{Alg. \ref{algo:bridge-update}}}\label{line: refertoupdatebridge}
        }
    }
    \Return{$G$}
\end{algorithm}

\subsubsection{Global Graph via Cross-Cluster Bridges}
\label{sec: globalAPG} 
\rtwo{Cluster-level graphs offer good local navigability, but cluster-by-cluster traversal is inefficient and duplicating sets across clusters introduces redundancy. This motivates weaving cluster-level graphs into a single, globally connected graph via bridge sets.}
As shown in Algorithm \ref{algo:bridge-update}, for a bridge set $P$, we gather both the neighbors already connected to $P$ from earlier clusters and the new neighbors identified in the current cluster (Line \ref{line: edgeold}-\ref{line: edgenew}).  If the total number of neighbors is within the degree limit $M$, we retain them all (Line \ref{line: remainall}). Otherwise, only top-$M$ closest neighbors are retained (Line \ref{line: remainM}). To ensure $P$ preserves its role as bridges, we enforce that at least one neighbor from each cluster in \( \mathcal{C}_{\text{top}}(P) \) is included in the final neighbor set (Line \ref{line: atleast1}-\ref{line: atleast2}). For example, if P belongs to clusters $C'_2$ and $C'_4$, this step ensures its final neighbor list will contain at least one neighbor from $C'_2$ and one from $C'_4$.  This prevents newly added connections from overwriting prior inter-cluster links and preserves global connectivity. Finally, the neighborhood of $P$ is updated to reflect the new connection set (Line \ref{line: updateneighbor}).
Note that each \( P \) corresponds to a \emph{unique} vertex in the global graph: logically assigned to multiple clusters, but physically stored as a single shared node. This design maintains intra-cluster structural coherence, supports efficient cross-cluster traversal via inter-cluster bridges, and eliminates redundant vertex duplication.

\begin{example}
    Figure \ref{fig:globalgraph} further illustrates the graph construction process. In Figure \ref{fig:globalgraph}(a), $P_3$ is assigned to cluster $C'_1$ based on its informative vectors (solid-edge circles). Within the cluster, a local graph is constructed using qEMD-based similarity. Since $P_3$ also belongs to another cluster $C'_2$, it becomes a natural bridge between clusters, as shown in Figure \ref{fig:globalgraph}(b). Figure~\ref{fig:globalgraph}(c) presents a complete toy example of the resulting dual-graph structure.
\end{example}

\begin{algorithm}[t]
    \caption{Update Bridges}
    \label{algo:bridge-update}
    \LinesNumbered
    \small
    \KwIn{Set $P$, new neighbors $C_{\text{new}}$, $G = (V, E)$, $\mathcal{C}_\text{top}(P)$, $M$}
    \KwOut{Updated global graph $G$}

    $C_{\text{old}} \gets$ existing neighbors of $P$ in $E$\; \label{line: edgeold}
    $C_{\text{all}} \gets C_{\text{new}} \cup C_{\text{old}}$\; \label{line: edgenew}

    \If{$|C_{\text{all}}| \le M$}{
        $C_{\text{final}} \gets C_{\text{all}}$\; \label{line: remainall}
    }
    \Else{
        $C_{\text{final}} \gets$ $M$ closest vertices of $P$ in $C_{\text{all}}$ by qEMD\; \label{line: remainM}

        \ForEach{$C_i \in \mathcal{C}_\text{top}(P)$\label{line: atleast1}}{
            \If{$C_{\text{final}} \cap C_i = \emptyset$}{
                $S_i \gets \{P \in C_{\text{all}} \mid P \in C_i\}$\;
                Replace farthest in $C_{\text{final}}$ with closest node from $S_i$\; \label{line: atleast2}

            }
        }
    }

    Replace $E[P]$ with edges $(P, P')$ for all $P' \in C_{\text{final}}$\; \label{line: updateneighbor}

    \Return{$G$}
\end{algorithm}

\begin{algorithm}[t]
    \caption{Inject Shortcuts}
    \label{algo:inject-shortcuts}
    \LinesNumbered
    \small
    \KwIn{Graph $G=(V,E)$, training pairs $\mathcal{T}$, $M$}
    \KwOut{Augmented GEM graph $G$}

    \ForEach{$(Q, P) \in \mathcal{T}$}{
        $C \gets$ $f'$ ANN results for Q in $G$\;  \label{line: f'ANNs}
        
        \If{$PID \notin C$ and both $P_{top}, P $ have degree $\leq M$ \label{line: multi-ifs}}{
            $P_{top} \gets$ 1-ANN in $C$\;
            Add a undirected edge $(P_{top}, P)$ to $E$\; \label{line: addshortcuts}
        }
    }
    \Return{$G$}
\end{algorithm}

\subsection{Graph Enhancement }
\subsubsection{Inject Shortcuts}
\label{sec: shortcuts} 
\begin{figure}[t]
\vspace{-1ex}
  \centering
  \includegraphics[width=0.3\textwidth]{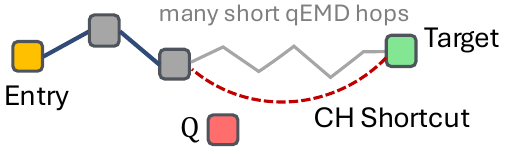}
  \caption{Graph Enhancement via Shortcuts}
  \label{fig:shortcut}
\end{figure}

We observed that some queries may require many short hops to reach a relevant result. This issue arises because, while EMD upper-bounds the Chamfer distance, the reverse does not hold. As a result, two vector sets may be close under Chamfer but remain distant in EMD space, preventing direct connections during graph construction and leading to longer traversal paths at query time.
To mitigate this, we propose a lightweight supervised graph augmentation technique that injects semantic shortcuts into the graph index. 
A shortcut is a direct edge connecting two vertices that are labeled as semantically close but might be distant in the graph. 
As detailed in Algorithm \ref{algo:inject-shortcuts}, we leverage training pairs \( (Q, P) \), where \( P \) is the \rcommon{human-annotated ground-truth to answer the query \( Q \)}, to augment the graph. 
\rcommon{Note that generating the training pairs does not require additional labeling effort or exhaustive search. In this work, we simply re-use  training sets that are widely used to train LLMs for generating
vector representations of data. For example, on MSMARCO dataset, to be introduced in Section \ref{sec:datasets}, provides \textit{“forms of training data, usually with one positive passage per training query,”} where a positive passage means \textit{“there is a direct human label that says the passage can be used to answer the query.”}  Here, we randomly sample from these existing training pairs to construct the shortcuts. }
For each pair, we run the search algorithm (Algorithm \ref{algo:query}) over the current graph to find the query's top-$f'$ ANNs (Line \ref{line: f'ANNs}). If \( P \) is not among them and both $P_{top}$ and $P$ have remaining degree capacity, we add an edge between them (Line \ref{line: multi-ifs}-\ref{line: addshortcuts}).
Since test queries are generally assumed to follow a similar distribution as the training data, such edges act as a shortcut, allowing future queries to reach the correct region more directly.
As illustrated in Figure~\ref{fig:shortcut}, the shortcut helps the search bypass long paths and suboptimal results. 
By adding EMD-based structural connectivity with Chamfer-based semantic alignment, we enhance graph reachability, search efficiency, and accuracy.

\subsubsection{Adaptive Cluster Cutoff}
\label{sec: adaptivek}

As discussed in Section \ref{sec: tfidfpruning}, a fixed cut-off threshold $r$ is suboptimal. 
To address this, we introduce an adaptive technique that employs a lightweight decision tree classifier \cite{scikit-learn} to predict an optimal per-set $r$. 
Our key idea is to retain the minimum cluster representation required for search-time discoverability.
Specifically, we first generate $r_{label}$ for each training pair $(Q, P)$. The label is defined as the rank of the first cluster in the TF-IDF-sorted profile $\mathcal{S}(P)$ that intersects with $Q$'s relevant cluster set $\mathcal{C}_\text{query}(Q)$. If no intersection occurs within the top-$r_\text{max}$ (default 10), the label is set to $r_\text{max}$.
Then, the decision tree is trained with the top-$r_\text{max}$ TF-IDF scores (padded if necessary) and the number of vectors in $P$ as input,  and learn to predict the number of clusters $r$ to retain. During the indexing pipeline, the trained model infers per-set $r$, which is then used for final cluster assignments.  This strategy effectively reduces the number of redundant assignments while preserving the necessary semantic coverage for effective retrieval.

\subsection{Query Processing}
\label{sec: query}

\begin{algorithm}[t]
\caption{Cluster-Guided Beam Search}
\label{algo:query}
\LinesNumbered
\small
\KwIn{Query $Q$, $G$, $\mathcal{C}_\text{quant}$, $\mathcal{C}_\text{index}$, parameters $t, ef, k$}
\KwOut{$k$ ANNs to $Q$}

Compute relevance scores $S_{c,q} = C \cdot Q^\top$\; \label{line: prefilter1}

$\mathcal{C}_{\text{query}} \gets$ union of top-$t$ centroids for each $q_i \in Q$\; \label{line: prefilter2}

$E \gets$ one random node from each $C_j \in \mathcal{C}_{\text{query}}$\; \label{line: multientry}

Initialize a shared result heap $\mathcal{R} \gets E$, a shared visited set $\mathcal{V} \gets E$\; \label{line: searchstart}
Initialize a local queue $\mathcal{W}_{ep}$ for each $ep \in E$\; \label{line: initW}
\While{any $\mathcal{W}_{ep}$ is not empty}{
    \ForEach{non-empty $\mathcal{W}_{ep}$ \textbf{in parallel} \label{line: until}}{
        $P \gets$ pop the element in $\mathcal{W}_{ep}i$ cloest to $Q$ by \text{qCH}\; \label{line: pop}
        $\tau \gets $ current furthest distance in $\mathcal{R}$\; \label{line: deftau}
        \If{\text{qCH}$(Q, P) > \tau$}{mark $\mathcal{W}_{ep}$ as empty\; \textbf{continue}\;\label{line: continue}}

        \ForEach{$P' \in \text{Neighbors}(P)$}{
            \If{$P' \notin \mathcal{V}$ and $\mathcal{C}_\text{top}(P') \cap \mathcal{C}_{\text{query}} \neq \emptyset$ \label{line: prune}}{
            compute $\text{qCH}(Q, P')$\; \label{line: pass1}
            insert $P'$ into $\mathcal{W}_{ep}$, $\mathcal{V}$, and $\mathcal{R}$\; \label{line: pass2}
            \If{$|\mathcal{R}| > ef$\label{line: maxef}}{Remove the furthest from $R$\; \label{line: maxef2}}
            }
            update $\tau$\; \label{line: updatetau}
        }
    }
}
$\mathcal{R} \gets$ rerank top-$k$ results in $\mathcal{R}$ by \text{CH}\;
\Return{$\mathcal{R}$} \label{line: searchend}
\end{algorithm}

Building on our index structure, we propose a cluster-guided beam search algorithm tailored for efficient multi-vector retrieval. The complete search procedure is detailed in Algorithm~\ref{algo:query}. At a high level, the algorithm improves upon traditional single-vector greedy search from three complementary perspectives: (1) coarse-grained cluster filtering, (2) informative multi-entry point initialization, and (3) cluster-aware early pruning during graph traversal.

\subsubsection{Cluster Filtering}
In multi-vector retrieval, distance computations are significantly more expensive than in the single-vector setting. To reduce this cost, it is crucial to eliminate highly unlikely vector sets as early as possible—ideally before any graph traversal begins. For this purpose, we draw inspiration from the initial filtering phase of ColBERTv2, which is extremely fast for coarse candidate selection. 
While this strategy alone may yield a large number of false positives, our downstream graph traversal module can refine these results with high precision.
This rationale leads us to retain ColBERTv2’s initial filtering strategy to generate coarse-grained cluster candidates. Specifically, we compute a query–centroid relevance matrix:
\begin{equation}
\small
    S_{c,q} = C \cdot Q^\top,
\end{equation}
where \( C \in \mathbb{R}^{k_2 \times d} \) is the centroid matrix and \( Q \in \mathbb{R}^{m_q \times d} \) is the query token matrix. For each query token \( q_i \), we select its top-$t$ nearest centroids and take the union across all tokens to form a relevant cluster set \( \mathcal{C}_{\text{query}} \subseteq \mathcal{C} \). This early pruning step is shown in Line \ref{line: prefilter1}-\ref{line: prefilter2} in Algorithm \ref{algo:query} and visually illustrated in Figure \ref{fig:searchGem}, where irrelevant clusters (in gray) are filtered out before the graph traversal.


\subsubsection{Multi-Entry Point Initialization}
\rtwo{Since relevant sets under the Chamfer distance may be distributed across multiple clusters,  relying on a single entry point can lead to biased initialization and longer traversal paths. Therefore, }
after identifying the relevant clusters, we initialize the search from multiple entry points, one from each cluster in \( \mathcal{C}_{\text{query}} \) (Line \ref{line: multientry}). As shown in Figure \ref{fig:searchGem}, two yellow vertices (ep1 and ep2) from two different clusters are randomly selected as the initial entry points.  
This multi-entry strategy, combined with the semantic shortcuts on the index side, effectively mitigates the fragmentation of Chamfer nearest neighbors and enhances overall retrieval effectiveness.

\subsubsection{Cluster-Guided Parallel Beam Search}
\label{sec:qCH}
Given the selected entry points, we perform a cluster-guided, multi-path parallel search over the graph \( G \) using quantized Chamfer distance (qCH). Similar to qEMD, qCH serves as a lightweight yet effective proxy for the expensive pairwise computation between vector sets, formalized as follows:
\begin{equation}
\small
    \text{qCH}(Q, P) = \sum_{q \in Q} \min_{p \in P} d_X(\text{NN}(q), \text{NN}(p)),
\end{equation}
where $\text{NN}(\cdot)$ is the nearest centroids in $\mathcal{C}_\text{quant}$. As elaborated before, although Chamfer is not used during index construction, it is upper-bounded by the EMD metric, leading that EMD-near neighbors are also reasonable candidates under Chamfer.
Line \ref{line: searchstart}- \ref{line: searchend} in Algorithm \ref{algo:query} details the query process. First, we initialize a global result heap $\mathcal{R}$, a global visited set $\mathcal{V}$, and a local priority queue $\mathcal{W}_{ep}$ for each entry point $ep \in E$ (Line \ref{line: searchstart}-\ref{line: initW}). The search then proceeds in parallel from these entry points. Within each thread, we repeatedly pop the element $P$ with the lowest qCH to the query (Line \ref{line: pop}) until $\mathcal{W}_{ep}$ is empty (Line \ref{line: until}). We denote $\tau$ as the current furthest distance to $Q$ in $\mathcal{R}$. If $\text{qCH}(Q, P)$ is already worse than $\tau$, the current path is deemed unpromising and terminated (Line \ref{line: deftau}-\ref{line: continue}). Otherwise, we continue expanding $P$'s neighbors. To 
boost query efficiency, we apply a cluster-aware pruning condition (Line \ref{line: prune}). That is, if $P'$ would pull the search into an irrelevant cluster (one not in $\mathcal{C}_\text{query}$), it is directly filtered out. If the $P'$ passes this filter, the qCH score is computed, and the local queue $\mathcal{W}_{ep}$, result heap $\mathcal{R}$, and visited set $\mathcal{V}$ are updated accordingly (Line \ref{line: pass1}-\ref{line: pass2}). $\mathcal{R}$ maintains a maximum of  $ef$ candidates, evicting the farthest node when necessary (Line \ref{line: maxef}-\ref{line: maxef2}). Note that both $\mathcal{R}$ and $\mathcal{V}$ are globally shared across threads, enabling implicit communication. If one path discovers a closer node, it updates the $\tau$ (Line \ref{line: updatetau}), which allows other paths to prune their search space earlier. Similarly, once a vertex is visited by any thread, it is skipped by all others, preventing redundant computation.
The search terminates when all $\mathcal{W}_{ep}$ queues are exhausted. In the final step, the top-$k$ candidates in \( \mathcal{R} \) are re-ranked using exact Chamfer distance before being returned.

\begin{figure}[t]
  \centering
  \includegraphics[width=0.4\textwidth]{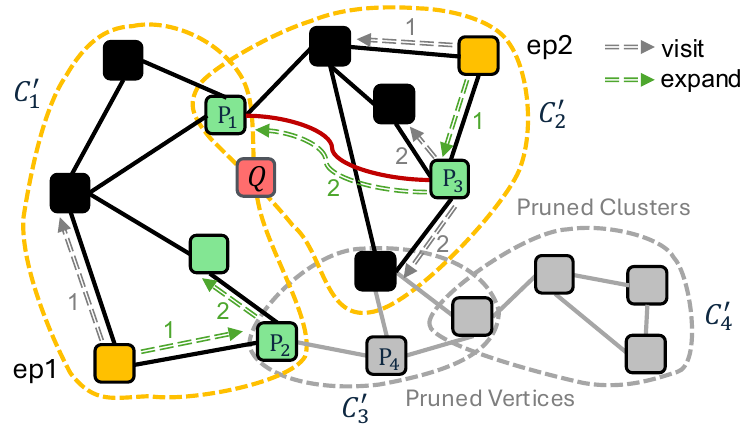}
  \caption{Search in GEM}
  \vspace{-2ex}
  \label{fig:searchGem}
\end{figure}
\begin{example}
    Figure~\ref{fig:searchGem} gives an example of the GEM search process for 1-ANN with \( ef = 2 \).  \( P_1 \) is the ground-truth nearest neighbor. Assume that the cluster filtering stage identifies two relevant clusters, \( \mathcal{C}_\text{query} = \{ C'_1, C'_2 \} \), while pruning $C'_3$ and $C'_4$. The search then begins in parallel from two entry points: ep1 and ep2 (yellow squares). In the first iteration (labeled with 1), the path of ep1 expands to \( P_2 \), while ep2 discovers \( P_3 \). These two candidates are inserted into the shared result heap \( \mathcal{R} \). In the second iteration (labeled with 2),  the path from ep1 continues by expanding from \( P_2 \).  When considering its neighbors,  \( P_4 \) is immediately pruned without a distance calculation because it resides in a pruned cluster. Concurrently, the other path expanding from \( P_3 \) jumps directly to \( P_1 \) via a semantic shortcut (highlighted in red), bypassing what would otherwise have been a multi-hop traversal.  
\end{example}

\subsection{Index Maintenance}

As a graph-based method, GEM naturally supports dynamic index maintenance, including both insertion and deletion of data points. 
To insert a new vector set into the current index, GEM first maps its vectors to the fine-to-coarse centroid hierarchy and prunes uninformative clusters using the TF-IDF–guided strategy, as described in Section \ref{sec:cluster}. Then, the set is incorporated into the graph by linking it to its nearest neighbors under the qEMD metric,   followed by updating the cross-cluster bridges to maintain both local proximity and global connectivity, as detailed in Section~\ref{sec:graphconstruction}.  
For deletions, we follow the lazy deletion strategy \cite{DBLP:journals/pami/MalkovY20}, which is commonly used in graph-based indexes: a vertex to be removed is simply marked as inactive, so it is skipped during search and can be physically pruned in later maintenance passes.







%% file: sec-experiment.tex
\section{Experiments}\label{sec:exp}

In this section, we conduct extensive experiments on real-world datasets to evaluate our GEM. We implement GEM\footnote{https://github.com/sigmod26gem/sigmod26gem} in C++ compiled with g++ using -Ofast optimization and openMP for parallelism. All experiments are run on a Ubuntu server with 4 Intel(R) Xeon(R) Gold 6218 CPUs (160 threads) and 1.5 TB RAM. 

\subsection{Experimental Settings}
\label{subsec:settings}
\subsubsection{\textbf{Datasets.}}
\label{sec:datasets}

\begin{table}[t]
\small
    \renewcommand\arraystretch{1.1}
	\caption{\rcommon{Summary of Datasets}}
	\label{tab:datasets}
	\small
	\begin{tabular}{c|c|c|c|c}
		\hline   
		{Datasets} & {\# Corpus} & {\# Vectors} & {\# Queries}  & {Modality} \\ 
		\hline 
		\rcommon{MS MARCO v1 \footnotemark[3]} & 8.8M & 597M & 6980 & Text-only\\
            \rcommon{LoTTe pooled \footnotemark[3]} & 2.4M & 339M & 2931 & Text-only\\
		\rcommon{OK-VQA \footnotemark[3]} & 114K & 14M & 5046 & Text + Image \\
           \rcommon{EVQA \footnotemark[3]} & 50K & 9.7M & 3750 & Text + Image \\
		\hline
	\end{tabular}
\end{table}
\footnotetext[3]{\rcommon{\href{https://github.com/microsoft/msmarco/tree/master}{MSMARCO weblink}, \href{https://ir-datasets.com/lotte.html}{LoTTE webLink}, \href{https://okvqa.allenai.org/download.html}{OK-VQA weblink}, \href{https://github.com/google-research/google-research/tree/master/encyclopedic_vqa}{EVQA weblink}}}
We evaluate GEM's performance on four widely-used multi-vector retrieval benchmarks, which are detailed in Table \ref{tab:datasets}. 
\textsf{MS MARCO v1} is widely used for in-domain evaluation with a retriever trained specifically for this task, while \textsf{LoTTE} serves as an out-of-domain benchmark. 
To further evaluate performance in complex multi-modal settings, we include two vision-language retrieval datasets, \textsf{OK-VQA} and \textsf{EVQA}, both involving queries with textual and visual content. 
\rcommon{
All four datasets provide training and test splits, in which each query is paired with one or more human-annotated relevant document IDs.  The training instances are used to enhance the graph, while queries from the test split are used to evaluate retrieval performance, with the corresponding relevance annotations treated as ground truth.}
For \textsf{MS MARCO}  and \textsf{LoTTE}, we employ ColBERTv2 \cite{DBLP:conf/naacl/SanthanamKSPZ22} to encode queries and documents into sets of vectors. \rtwo{Following prior work \cite{PLAID, DESSERT, MUVERA, IGP}, the dimensionality of each individual vector is set to $d=128$ by default.}
For multi-modal datasets, we adopt preFLMR \cite{DBLP:conf/acl/LinMCB24}, a fine-grained late-interaction retriever that jointly encodes text and image content into multi-vector representations ($d=128$). 
Both ColBERTv2 and preFLMR adopt the late interaction paradigm with set-to-set similarity computed via Chamfer distance, and have shown strong performance in text-only and multi-modal retrieval, respectively.

\subsubsection{\textbf{Competitors.}}

We compare GEM with four representative methods in multi-vector retrieval, as mentioned in Section \ref{sec: relatedwork}: PLAID \cite{PLAID}, DESSERT \cite{DESSERT}, MUVERA \cite{MUVERA} and IGP \cite{IGP}. We also include a graph-based baseline as discussed in Section \ref{sec:naivesolution},  denoted as the multi-vector graph (MVG). To ensure basic competitiveness, here we use qCH for both indexing and search in our experiments.

\subsubsection{\textbf{Parameter Settings.}}
\label{sec:para}Parameter settings of
competitors follow the original papers or their source codes. 
For PLAID, we set the number of centroids to $262K$ for MSMARCO and LoTTE, $32K$ for OKVQA and EVQA. 
All vectors are quantized to 2 bits per dimension for MS MARCO and LoTTE, and to 8 bits for OKVQA and EVQA. For DESSERT, {we set $C = 7$, $L = 64$ for MSMARCO, LoTTE, and set $C = 4$, $L = 768$ for OKVQA, EVQA. For MUVERA, we adopt $\text{R}_{reps} = 20$, $\text{K}_{sim} = 5$, and $\text{D}_{proj} = 32$ to generate single-vector representations, which are then compressed using PQ-256x8 and indexed with HNSW ($M=24$, $\texttt{ef\_construct}=80$).
For IGP, we use $262K$ centroids for MSMARCO and LoTTE, $32K$ for OKVQA and EVQA, with graph construction parameters set to $M = 24$, $\texttt{ef\_construct} = 200$.
For our GEM,  \rcommon{the number of codebooks $|\mathcal{C}_{\text{quant}}|$ follows the same empirical formula from prior studies \cite{PLAID,DBLP:conf/naacl/SanthanamKSPZ22}, \ie the nearest power of two to $16 \times \sqrt{\#\text{vectors}}$. This yields $|\mathcal{C}_{\text{quant}}|
= 2^{\left\lfloor \log_{2}\!\left(16\sqrt{597M}\right) \right\rfloor} \approx 262k$ for MASMARCO. Similarly,  $|\mathcal{C}_\text{quant}| = 262K$ for LoTTE, and $32K$ for OK-VQA and EVQA.} \rcommon{The number of clusters $|\mathcal{C}_{\text{index}}|$  is scaled with dataset size to balance granularity and indexing cost. Specifically, $|\mathcal{C}_{\text{index}}|$ is set to $40K$ for MSMARCO, $10K$ for LoTTE, and $1K$ for OK-VQA and EVQA, so that each cluster contains about $10^{3}\!-\!10^{4}$ objects.
The graph construction parameters $M = 24$ and \texttt{ef\_construction}$ = 80$ and} \rcommon{20\% of the training instances are randomly sampled to construct the shortcuts.}
At query time, we set $t = 4$ and vary \texttt{ef\_search}, $\#rerank$ to trade off efficiency and accuracy.
For MVG, we adopt the same indexing parameters as GEM. All competitors are also tuned to their best query performance. 








\subsubsection{\textbf{Evaluation Metrics.}}


\rthree{Following previous works \cite{PLAID, DESSERT, DBLP:conf/acl/LinMCB24}}, we adopt four metrics to evaluate query performance: \textit{Recall}, \textit{Mean Reciprocal Rank (MRR)}, \textit{Success}, and query latency; and two metrics to evaluate indexing performance: indexing time and index size.
Given a query set $\mathcal{Q}$, with ground-truth set $\mathcal{G}_Q$  for each query $Q \in \mathcal{Q}$, and top-$k$ retrieved results $\mathcal{R}_Q^{(k)}$, Recall@$k$ measures the proportion of relevant items retrieved: 
\begin{equation}
\small
    \text{R@}k = \frac{1}{|\mathcal{Q}|} \sum_{Q \in \mathcal{Q}} \frac{|\mathcal{G}_Q \cap \mathcal{R}_Q^{(k)}|}{|\mathcal{G}_Q|}. 
\end{equation}
MRR@$k$ captures the reciprocal rank of the first relevant item:
\begin{equation}
    \small
    \text{MRR}@k = \frac{1}{|Q|} \sum_{Q \in \mathcal{Q}} \frac{1}{\text{rank}_Q},
\end{equation}
\rtwo{Success@$k$ indicates whether any relevant item is present in the top-$k$, defined as follows:}
\rtwo{\begin{equation}
\small
    \text{S@}k = \frac{1}{|Q|} \sum_{Q \in \mathcal{Q}} \mathbb{I}[ \mathcal{G}_Q \cap \mathcal{R}_Q^{(k)} \neq \emptyset].
\end{equation}}
Query latency is reported as the average runtime over $10$ repeated runs for each dataset.
\begin{remark}
    \rcommon{Note that the ground-truth set $\mathcal{G}_Q$ in the above definitions refers to results with \textit{"a direct human label that says the passage can be used to answer the query"}, as stated in MSMARCO. Other datasets follow the same principle. Unlike ANN benchmarks, where the ground truth is typically defined as the exact $k$NN obtained by brute-force search in the embedding space, our results below directly reflect end-to-end semantic similarity retrieval performance.}
\end{remark}

\subsection{Performance Analysis}

\begin{table*}[t]
  \centering
  \caption{End-to-end Retrieval Performance Overview}
  \label{tab:main_results}
  \renewcommand{\arraystretch}{1.3}
  \setlength{\tabcolsep}{3.5pt}
  \footnotesize
  \begin{tabular}{l|cccc|cccc|cccc|cccc}
    \toprule
    \multirow{2}{*}{Method} 
    & \multicolumn{4}{c|}{MSMARCO} 
    & \multicolumn{4}{c|}{LOTTE} 
    & \multicolumn{4}{c|}{OKVQA} 
    & \multicolumn{4}{c}{EVQA} \\
    \cmidrule(r){2-5} \cmidrule(r){6-9} \cmidrule(r){10-13} \cmidrule(r){14-17}
    & R@100 & S@100 & MRR@10 & T (ms) 
    & R@100 & S@100 & MRR@10 & T (ms) 
    & R@100 & S@100 & MRR@10 & T (ms) 
    & R@100 & S@100 & MRR@10 & T (ms) \\
    \midrule
    PLAID          
    & 0.913 & 0.919 & 0.395 & 352 
    & 0.745 & 0.910 & 0.558 & 462 
    & 0.400 & 0.735 & 0.225 & 570 
    & 0.895 & 0.895 & 0.476 & 211 \\
    DESSERT        
    & 0.890 & 0.905 & 0.372 & 344 
    & 0.748 & 0.893 & 0.541 & 362 
    & 0.371 & 0.723 & 0.163 & 642
    & 0.837 & 0.837 & 0.370 & 552 \\
    MUVERA         
    & 0.902 & 0.915 & 0.385 & 310 
    & 0.736 & 0.901 & 0.547 & 441 
    & 0.369 & 0.723 & 0.219 & 605
    & 0.878 & 0.878 & 0.471 & 446 \\
    IGP        
    & 0.895 & 0.901 & 0.389 & 340
    & 0.731 & 0.899 & 0.553 & 338 
    & 0.402 & 0.732 & 0.215 & 549
    & 0.892 & 0.892 & 0.440 & 458 \\
    MVG         
    & 0.895 & 0.905 & 0.421 & 360 
    & 0.701 & 0.879 & 0.555 & 231 
    & 0.381 & 0.719 & 0.230 & 427 
    & 0.880 & 0.880 & 0.468 & 450 \\
    \midrule
    \textbf{GEM (Ours)}   
    & \textbf{0.915} & \textbf{0.920} & \textbf{0.447} & \textbf{140} 
    & \textbf{0.749} & \textbf{0.910} & \textbf{0.592} & \textbf{210} 
    & \textbf{0.410} & \textbf{0.754} & \textbf{0.236} & \textbf{165} 
    & \textbf{0.895} & \textbf{0.895} & \textbf{0.477} & \textbf{197} \\
    \bottomrule
  \end{tabular}
\end{table*}
\subsubsection{\textbf{Query Performance}}
In this section, we study the end-to-end retrieval performance of all methods under default settings, as summarized in Table~\ref{tab:main_results}, Table~\ref{tab:msmarco_lotte_results}, and Figure~\ref{fig:overall_performance}.

Table \ref{tab:main_results} provides an overview of recall, success, MRR, and end-to-end latency for all methods across all datasets. Clearly, GEM offers the best overall query performance on all datasets.  It adapts well to in-domain, out-of-domain, and more complex multi-modal datasets, outperforming all competitors on every query quality metric while maintaining the lowest latency.  In particular, GEM achieves speedups of up to 2.5$\times$ on MSMARCO, 2.2$\times$ on LoTTE, 3.9$\times$ on OKVQA, and 2.8$\times$ on EVQA, all with better query quality. \rthree{For EVQA, each query has exactly one annotated relevant document, thus $\mathcal{G}_Q \cap \mathcal{R}_Q^{(k)}$ can only be size $0$ or $1$. If the ground truth is in the retrieved top-$k$: $\text{recall} = 1/1=1, \text{success} = 1$; 
otherwise: $\text{recall} = \text{success} = 0$. Hence, its averaged {recall} and {success} reported in Table ~\ref{tab:main_results} and Figure \ref{fig:overall_performance} are always identical.}
These experimental results validate our design goal: preserving the semantic richness of multi-vector representations to achieve high accuracy, while leveraging an efficient graph index for speed. 
The reason GEM achieves the best performance can be concluded as follows: 
1) Compared with PLAID, GEM indexes at a set-level granularity, treating the entire vector set as the basic unit. This is fundamentally different from PLAID, which indexes all vectors individually. By doing so, GEM significantly reduces the number of false positive candidates returned due to token-level matches, avoiding the need for expensive refinement and boosting efficiency.
2) Compared with DESSERT, which is often bottlenecked by an exhaustive scan of all set sketches, GEM's graph structure provides a powerful pruning mechanism, enabling significantly faster navigation to the most promising candidates.
3) Compared with MUVERA, which suffers significant accuracy degradation by compressing a vector set into a single vector, GEM preserves set-level semantic nuances, thereby easily achieving superior query quality that MUVERA struggles to match even when its FDEs are tuned to an impractically high $20480$ dimensions.
4) Compared with IGP, which extends token-level indexing into a graph structure over quantized centroids for incremental candidate filtering, GEM constructs a set-level graph where each node represents a complete vector set.
This design captures inter-set relations that centroid-level graphs overlook, enabling faster and more accurate retrieval.
5) Compared with the naive MVG, while it preserves multi-vector inputs, its naive graph suffers from fragmented neighbors, local optima, and lacks effective distance approximation and pruning strategies.

\begin{figure*}[t]
  \centering
  \includegraphics[width=0.98\linewidth]{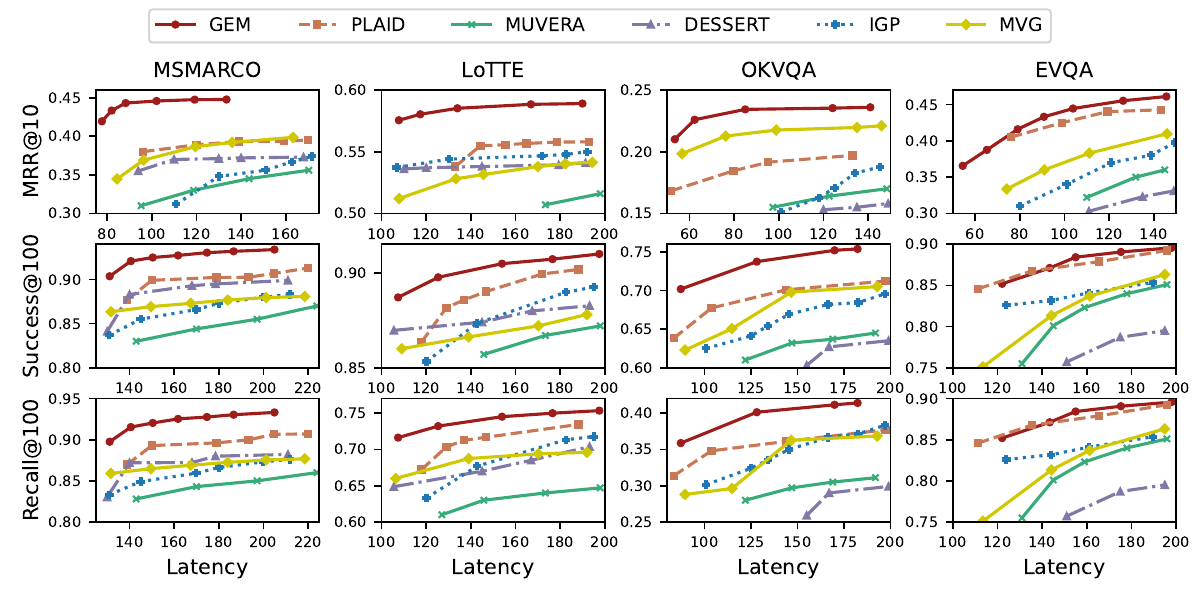}
  \vspace{-3ex}
  \caption{\rcommon{Accuracy–Latency Trade-off Overview}}
  \vspace{-1ex}
  \label{fig:overall_performance}
\end{figure*}

\begin{table}[t]
  \centering
  \caption{End-to-End Retrieval Perfomance Varying $k$}
  \vspace{-1ex}
  \label{tab:msmarco_lotte_results}
  \renewcommand{\arraystretch}{1.3}
  \setlength{\tabcolsep}{3.5pt}
  \footnotesize
  \begin{tabular}{l|ccc|ccc}
    \toprule
    \multirow{2}{*}{Method} 
    & \multicolumn{3}{c|}{MSMARCO} & \multicolumn{3}{c}{LoTTE} \\
    \cmidrule(r){2-4} \cmidrule(r){5-7}
    & R@10/T & R@100/T & R@1k/T 
    & S@10/T & S@100/T & S@1k/T \\
    \midrule
    PLAID          
    & 68.8/222 & 91.3/352 & 97.5/352 & 76.8/288 & 91.0/462 & 94.5/462 \\
    DESSERT        
    & 66.2/298 & 89.0/344 & 97.2/336 & 74.7/362 & 89.3/362 & 94.3/477 \\
    MUVERA         
    & 67.7/286 & 90.2/310 & 97.1/444 &  74.1/397  & 90.1/441  & 94.4/461 \\
    IGP
    & 66.9/279 & 89.5/340 & 96.9 /399 & 75.5/311 & 89.9/338 & 94.0/412 \\
    MVG         
    & 69.2/293 & 89.5/360 & 94.2/436 & 74.7/303 & 87.9/331 & 94.2/442 \\
    
    \midrule
    \textbf{GEM}   
    & \textbf{71.1}/\textbf{88} & \textbf{91.5}/\textbf{140} & \textbf{97.5}/\textbf{212} & \textbf{77.3}/\textbf{133} & \textbf{91.0}/\textbf{210} & \textbf{94.5}/\textbf{251} \\
    \bottomrule
  \end{tabular}
  \vspace{-3ex}
\end{table}
Table \ref{tab:msmarco_lotte_results} reports how query quality and latency change as the number of returned results $k$ increases. 
Due to space limitations, we only present results on two datasets. \rthree{Following our competitors \cite{PLAID, DESSERT}, we report recall on MSMARCO and success on LoTTE.}  \rcommon{On MSMARCO, GEM uses \texttt{ef\_search} = 800, 2000, 8000 for $k$ = 10, 100, 1k, respectively. On LoTTE, GEM uses \texttt{ef\_search} = 1000, 5000, 8000 for $k$ = 10, 100, 1k, respectively. 
} 
We make the following observations: 
(1) GEM consistently achieves the best query performance across all $k$ values, offering higher Recall@$k$ and Success@$k$ while maintaining significantly shorter query times. For example, on MSMARCO, GEM requires only $88$ ms, $140$ ms, and $212$ ms for $k=10,100,$ and $1000$ respectively, whereas the best baselines require $222$ ms (PLAID), $310$ ms (MUVERA), and $336$ ms (DESSERT) to reach comparable recall. On LoTTE, GEM completes in $133$ ms, $210$ ms, and $251$ ms, while the fastest baselines still require $288$ ms (PLAID), $338$ ms (IGP), and $412$ ms (IGP) to achieve similar quality.
\rtwo{(2) Our naive solution, MVG, can compete with several well-tuned baselines for small $k$. Though non-metric distance induces fragmented neighbors and local optima, this issue is not overly severe when the target $k$ is small. By slightly increasing \texttt{ef\_search}, greedy graph traversal  can escape local minima and reach to the ground truth eventually. As a result, MVG retains the key advantage of graph-based indexes, \ie it accesses fewer candidates than LSH-based (DESSERT) and IVF-style  (PLAID, IGP) methods to achieve the same recall. When further combined with our qCH quantization, which reduces the cost of per-candidate distance computation, MVG exhibits a favorable accuracy–latency trade-off.}
\rtwo{However, MVG does not consistently outperform other methods. For instance, on MSMARCO, other methods reach around 97\% recall when  $k=1000$, while MVG  saturates at $94.2\%$. Even with larger \texttt{ef\_search} and longer query time, further improvement is limited by the suboptimal graph structure resulting from the non-metric nature of Chamfer distance.
This underscores the necessity of metric decoupling and other optimizations, such as shortcuts and multi-entry search, in GEM for achieving stable performance gains.}
(3) As expected, in the out-of-domain setting (LoTTE), retrieval is more challenging than in-domain setting (MSMARCO), and all methods require more time to achieve good quality. GEM remains consistently more accurate and faster, often requiring only $1/2$ or even $1/3$ of the time of other methods.

Figure \ref{fig:overall_performance} provides a more comprehensive view of the trade-off between query accuracy and efficiency of all methods.
From the figures, we have the following observations: 
(1) GEM consistently achieves the best trade-off. Among all methods, GEM requires the least time to reach the same MRR, success, \rcommon{or recall},  indicating the most favorable balance between accuracy and efficiency. For the same MRR@10, GEM is up to 16x faster than DESSERT (on OKVQA), up to 10x faster than MUVERA (on OKVQA), up to 8x faster than IGP (on OKVQA), and up to 5x faster than PLAID (on MSMARCO).
(2) Visual-language retrieval is inherently more challenging. On OKVQA and EVQA, even with increased search time, Success@100 reaches only about 75–89\%, while in text-only QA tasks, our algorithm can easily exceed 90\%. Notably, this bottleneck is not on the retrieval side: even brute-force search on OKVQA and EVQA achieves only around 77–90\% Success@100. This points to the need for stronger embedding models and improved multimodal feature alignment mechanisms.
\rtwo{(3) MVG can outperform some competitors on certain datasets, such as complex multi-modal  OKVQA and EVQA datasets. This advantage can be attributed to its tailored set-level indexing structure, which preserves fine-grained semantics. In contrast, methods like MUVERA maps a multi-vector set to a single vector, inevitably causing non-trivial information loss. This loss worsens with data complexity, leading to degraded query performance.}
(4) GEM and PLAID are highly robust, maintaining stable performance across all datasets. PLAID is often the second-best method. In contrast, IGP, DESSERT,  and MUVERA exhibit substantial performance variation across datasets.
\begin{figure}[t]
    \centering
    \includegraphics[width=\linewidth]{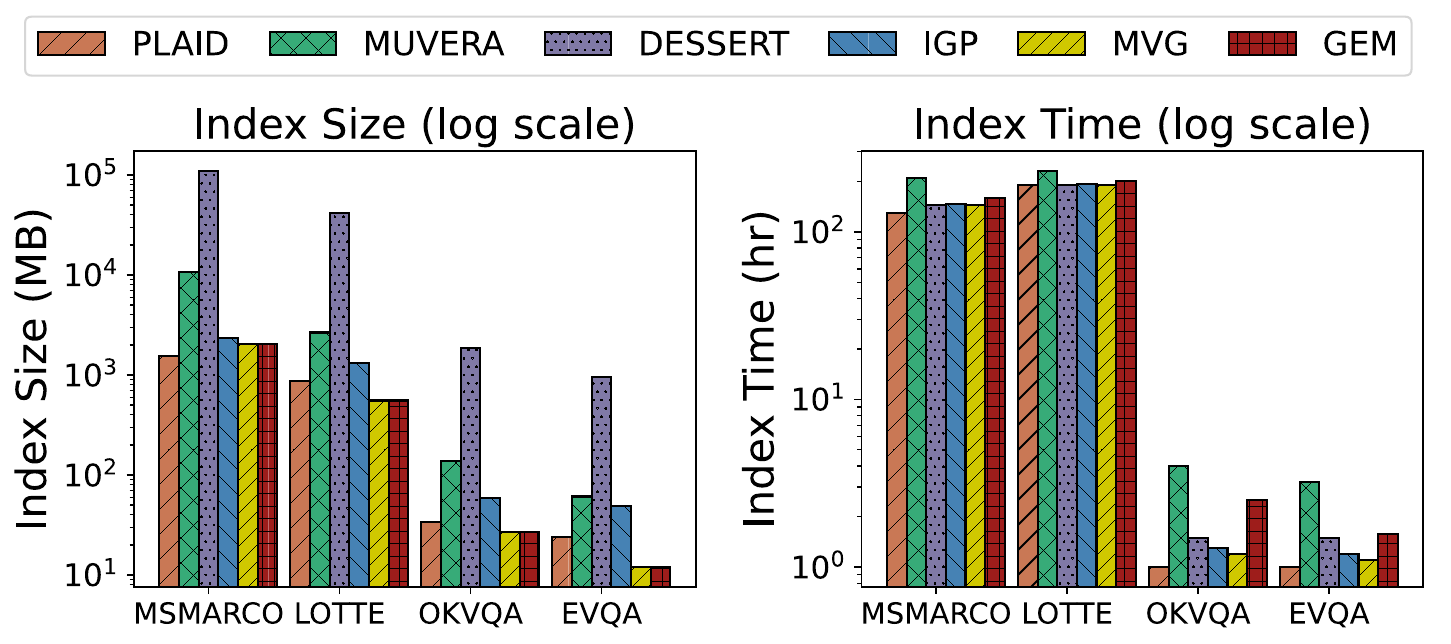}
    \caption{Indexing Performance on All Datasets}
    \vspace{-2ex}
    \label{fig:index}
\end{figure}
\subsubsection{\textbf{Indexing Performance}}
The indexing time and index size of all methods under default settings are summarized in Figure~\ref{fig:index}. Here, the index size includes only the index files, excluding raw data, to clearly highlight the differences among indexing methods. We make two key observations: (1) GEM’s index size is small (\eg 2 GB on MSMARCO, 27MB on OKVQA), significantly smaller than MUVERA (10.5 GB on MSMARCO, 139 MB on OKVQA) and DESSERT (107 GB on MSMARCO, 1.8GB on OKVQA). This compactness is attributed to the graph structure of GEM, which primarily stores edge information. This storage overhead is independent of whether the setting is under single- or multi-vector, and is negligible compared to the dataset size. In contrast, DESSERT requires storing a large number of hash values generated by LSH, while MUVERA needs to store Fixed Dimensional Encodings (FDEs) for all vector sets, which scale directly with the dataset size and make it less suited to large-scale scenarios. 
GEM’s index size is comparable to that of IGP (2.3 GB on MS MARCO, 59 MB on OKVQA) and PLAID (1.5 GB on MS MARCO, 34 MB on OKVQA), but GEM delivers a substantial advantage in both efficiency and accuracy. (2) On MSMARCO and LoTTE, all methods have comparable indexing time. This is because PLAID, DESSERT, IGP, and GEM all rely on a clustering step, which dominates the construction cost. MUVERA does not involve clustering but has to convert every vector set into a single vector, which is also time-consuming. Both clustering and vector conversion can be accelerated on GPUs, making this step less of a practical concern. On smaller datasets, OKVQA and EVQA, the clustering cost becomes less significant, and GEM’s graph construction is slower than PLAID’s IVF, IGP's single vector graph, and DESSERT’s LSH indexing. MUVERA, which requires both vector conversion and graph construction, remains the slowest. Overall, GEM achieves competitive indexing time and compact index size while providing superior retrieval performance.

\begin{figure}[t]
  \centering
  \includegraphics[width=\linewidth]{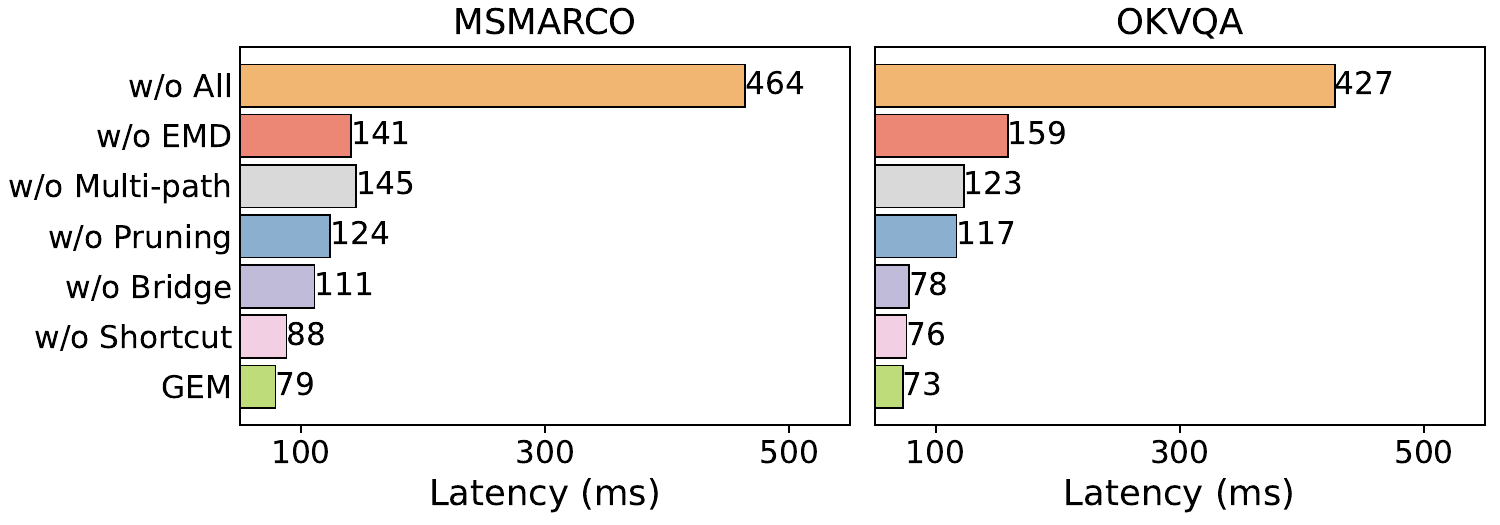}
  \caption{Ablation Study on the GEM's Components}
  \vspace{-2ex}
  \label{fig:albation_study}
\end{figure}

\subsection{Ablation Study}

In this section, we evaluate how much each component of GEM contributes to the overall performance. 
\rtwo{We disable key modules individually and compare the query latency required to reach comparable retrieval accuracy. Due to space constraints, we only present two representative scenarios, \ie MRR@10 = 0.42 on MSMARCO, and MRR@10 = 0.30 on OKVQA.} As shown in Figure \ref{fig:albation_study}, removing different components results in varying degrees of latency increase. 



\subsubsection{\textbf{w/o EMD distance.}} We replace the qEMD with qCH for graph construction, which doubles the latency on both datasets. The absence of metric properties degrades the graph quality, confirming the importance of the decoupling strategy in GEM.

\subsubsection{\textbf{w/o multi-path strategy.}} We disable multi-path search from multiple entry points, instead enqueuing all entry points into a single search queue $\mathcal{W}$ and expanding greedily. This change increases latency by 70–80\%. 
Without synchronized paths sharing the visited and best-found result sets, pruning is less aggressive and the search space contracts more slowly. 
To reach the correct results, a larger \texttt{ef\_search} is needed, which increases latency. These results confirm that multi-entry search accelerates convergence.

\subsubsection{\textbf{\rthree{w/o adaptive TF-IDF cluster pruning.}}} 
We disable the adaptive adjustment of clusters per document and fix $|\mathcal{C}_\text{top}| = 3$ for all documents\rthree{--a value nearly identical to the average $|\mathcal{C}_\text{top}|$ on MSMARCO and OKVQA when TF-IDF is enabled (Section \ref{sec: tfidfpruning}). This setup isolates the effect of cluster granularity and allows a direct, fair comparison  of \textit{adaptive v.s. fixed} $\mathcal{C}_\text{top}$ values.
As observed, performance degrades} because a fixed threshold hurts recall for long, topically diverse documents, while short documents suffer efficiency loss from retaining unnecessary clusters. This result highlights the importance of dynamic TF-IDF–guided pruning in balancing accuracy and efficiency.

\subsubsection{\textbf{w/o bridge constraint.}} In this variant, when the degree of a bridge $P$ exceeds the limit, we retain only its $M$ closest neighbors, rather than enforcing it to keep at least one neighbor from each cluster in $\mathcal{C}_\text{top}(P)$.
This change greatly degrades performance, with the drop on MSMARCO more pronounced than on OKVQA due to the larger corpus size. This result shows the importance of bridges in maintaining graph connectivity, especially on large datasets.


\subsubsection{\textbf{w/o shortcuts enhancement.}} We remove shortcuts that directly connect sets that may be close under Chamfer but require many hops to reach over the EMD graph. This leads to a modest degradation in query latency. While the co-designed architecture of GEM is already robust for most queries, the shortcut mechanism is a useful addition for handling a few hard-to-find cases.

It is worth noting that these components are co-designed to work synergistically. The optimal performance of GEM relies on their combined effect. Removing all of them (w/o all) leads to a severe degradation, with latency increasing by 5–6x.

\subsection{Further Analysis}

\begin{figure}[t]
  \centering
  \includegraphics[width=\linewidth]{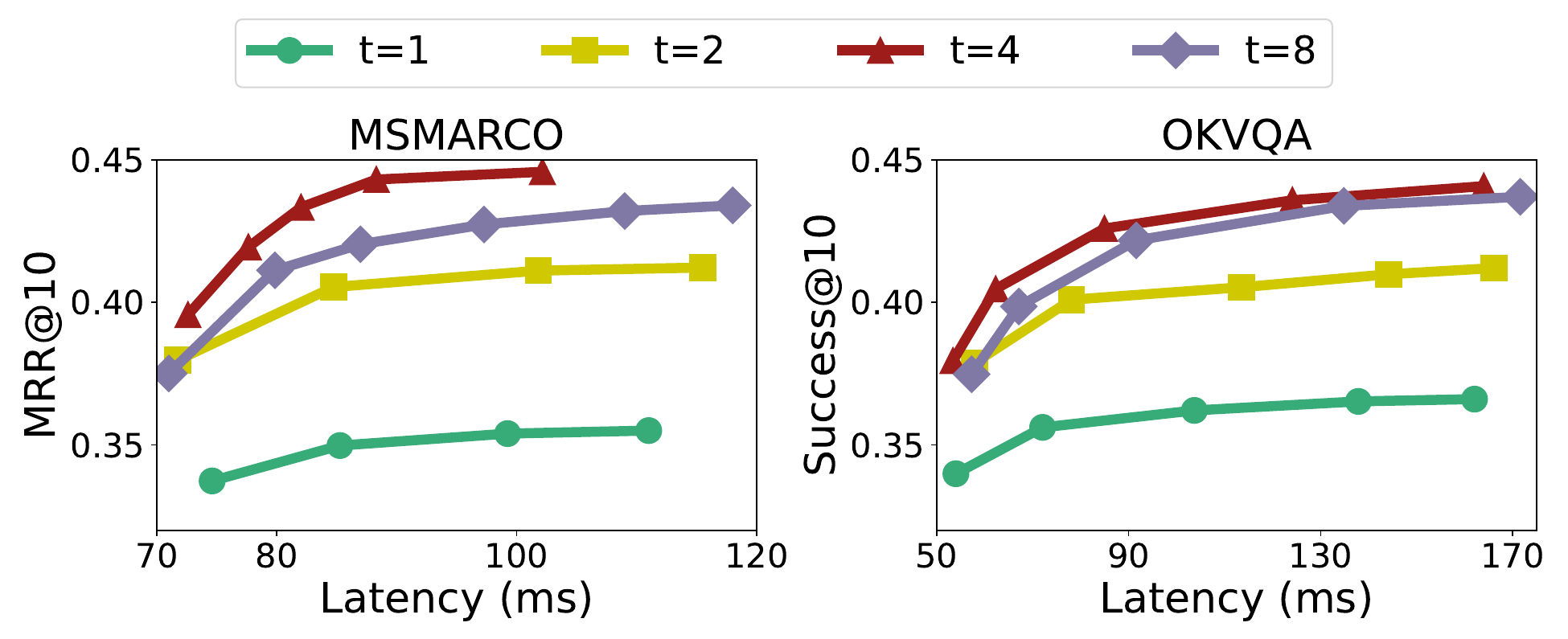}
  \caption{Impact of \bm{$t$} on Retrieval Performance}
  \label{fig:nprob_latency_accuracy}
\end{figure}

\begin{figure}[t]
  \centering
  \includegraphics[width=\linewidth]{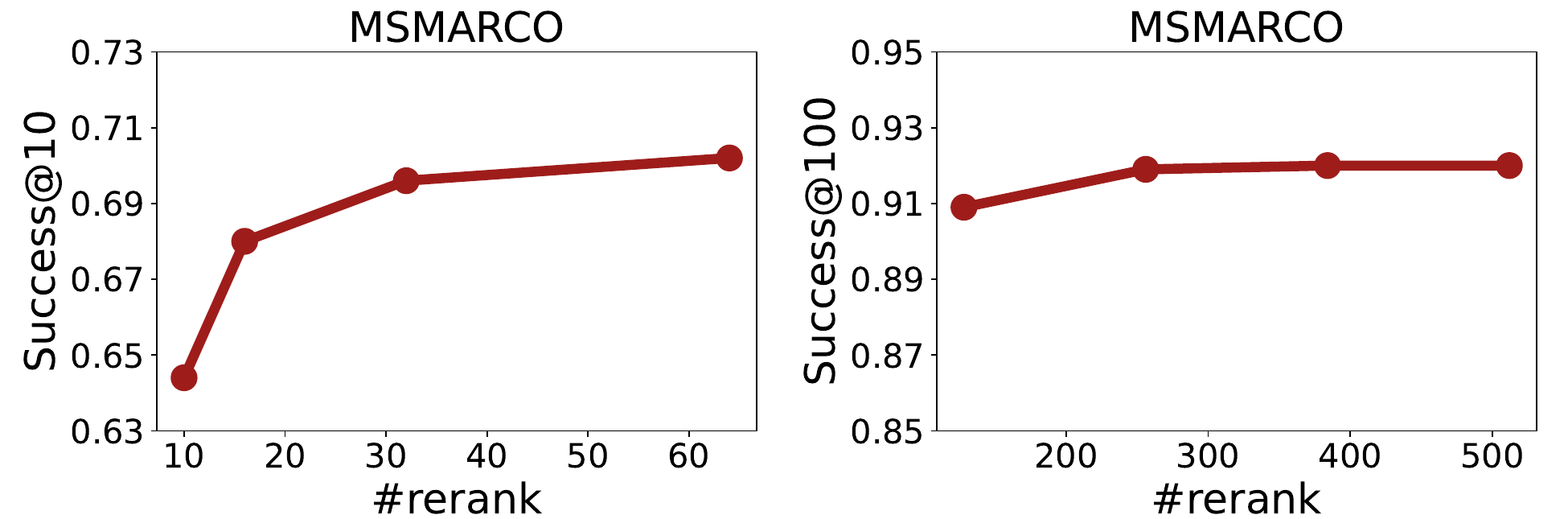}
  \caption{Impact of \#rerank on Retrieval Performance}
  \vspace{-2ex}
  \label{fig:rerank_effect}
\end{figure}
\subsubsection{\textbf{Effect of $t$.}}
In this experiment, we study the impact of parameter $t$, which controls the number of nearest clusters each $q_i \in Q$ identifies in the early pruning step. We vary $t \in \{1, 2, 4, 8\}$ and report results on MSMARCO and OKVQA for brevity. Larger $t$ makes the pruning more conservative: pruning fewer clusters to potentially increase recall, but at the cost of higher latency. As shown in Figure~\ref{fig:nprob_latency_accuracy}, both datasets exhibit similar trends.
When $t=1$, pruning is too aggressive, leading to insufficient accuracy. Performance improves substantially at $t=2$,  while the gains begin to saturate as $t$ increases further to $8$. Therefore, we set $t=4$ by default as it provides a favorable balance between accuracy and efficiency.

\subsubsection{\textbf{Effect of \#rerank.}}

In this experiment, we evaluate the impact of \textit{rerank\_k}. This parameter determines the size of the candidate pool retrieved from the graph search, which is then re-ranked using the exact Chamfer distance to obtain the final top-k results. Figure~\ref{fig:rerank_effect} reports results for $k=10$ and $100$. For $k=10$, the choice of \textit{rerank\_k} has a clear effect on accuracy: a larger \textit{rerank\_k} increases the likelihood that the true best results are included and ranked near the top.  For $k=100$, increasing \textit{rerank\_k} has a more modest impact. This is because our algorithm is already highly effective, with a strong probability of placing the true results within its  top-100 candidates, leaving limited room for further improvement.




\subsubsection{\textbf{Effect of Index Parameters.}}
In this experiment, we investigate the impact of two key indexing parameters: the maximum number of neighbors per vertex ($M$) and the size of the candidate list during construction (\texttt{ef\_construction}). As shown in Figure~\ref{fig:search_performance_vs_latency} on MSMARCO, with $M=8$ and \texttt{ef\_construction} = 24, the index is small and fast to build, but the graph quality is insufficient, resulting in lower query accuracy. Increasing to $M = 48$ and \texttt{ef\_construction} = 200 significantly enlarges the index size and construction time, but query performance improves only marginally. This is because each vertex connects to too many neighbors, increasing the number of candidates to verify at each hop and thereby hurting latency. Our default configuration of $M=24$ and \texttt{ef\_construction} = 80 achieves the best overall trade-off, offering competitive indexing performance and superior query efficiency and accuracy.

\begin{figure}[t]
  \centering
  \includegraphics[width=\linewidth]{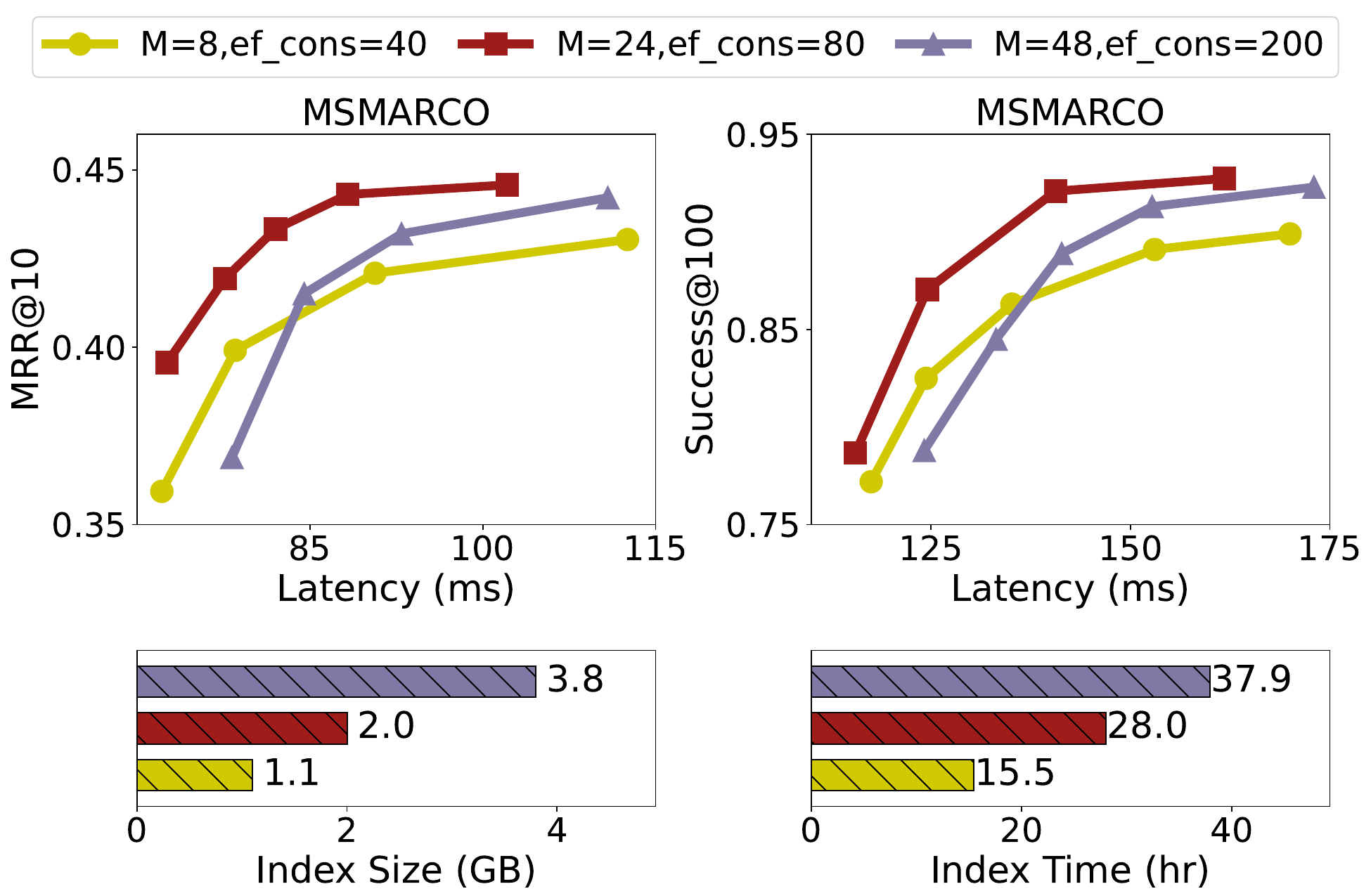}
  \caption{Impact of \bm{$M$} and $\texttt{ef\_construction}$}
  \label{fig:search_performance_vs_latency}
\end{figure}

\subsubsection{\textbf{Effect of $N$ and $m$.}} 

\begin{figure}[t]
\centering
\setlength{\abovecaptionskip}{5pt}
\includegraphics[width=\linewidth]{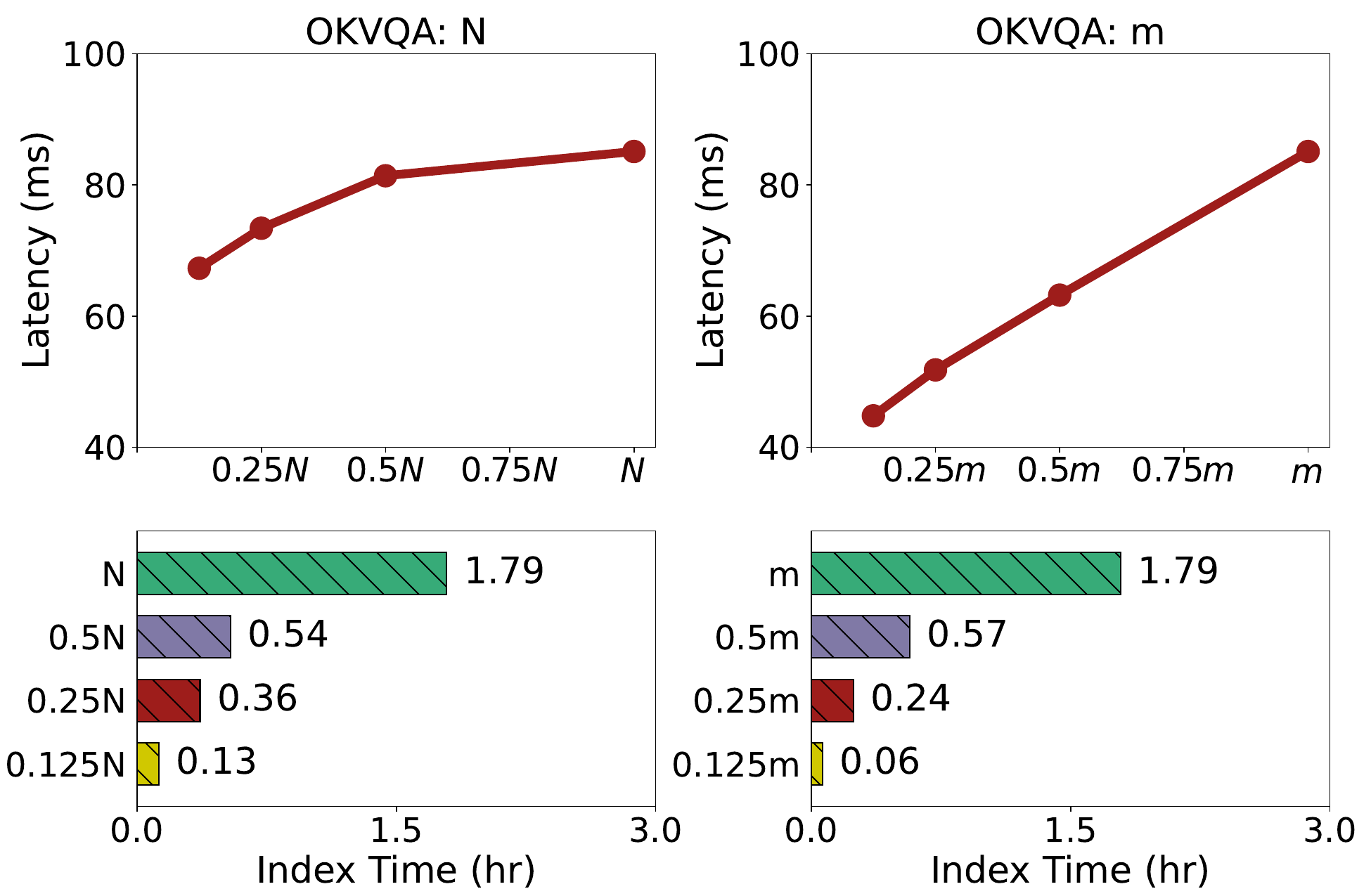}
\caption{Impact of corpus size $N$ and vector set size $m$}
\label{fig:scalability}
\end{figure}

In this experiment, we evaluate scalability with respect to the corpus size ($N$) and the average number of vectors per set ($m$). 
We randomly sample $\tfrac{1}{8}N$, $\tfrac{1}{4}N$, $\tfrac{1}{2}N$, and $N$ vector sets from the original corpus, and randomly sample $\tfrac{1}{8}m$, $\tfrac{1}{4}m$, $\tfrac{1}{2}m$, and $m$ vectors from each set to form new subsets. We vary one factor at a time.
Figure~\ref{fig:scalability} reports the effects of these variations on both query latency and index construction time.
Increasing $N$ causes a slight rise in latency, \eg from $67$ ms ($\frac{1}{8}N$) to $85$ ms ($N$), consequently increasing indexing time (0.13 h to 1.79 h) due to the search-based construction procedure.
This trend aligns with the logarithmic query cost of graph-based indexes, demonstrating GEM's strong scalability across different corpus scales. Increasing $m$ exerts a much stronger impact on efficiency, \eg query latency rises from $44$ ms ($\frac{1}{8}m$) to $85$ ms ($m$) and indexing time from $0.06~\text{h}$ to $1.79$ h. 
The higher sensitivity to $m$ arises from the set-to-set distance calculations. 
This also explains why many studies focus on optimizing the embedding training itself to preserve semantics more effectively with fewer vectors. 
As discussed in Section \ref{sec: relatedwork}, our method can be seamlessly integrated with such advances.

\begin{figure}[t]
  \centering
  \begin{minipage}[t]{0.235\textwidth}
    \centering
    \includegraphics[width=\textwidth]{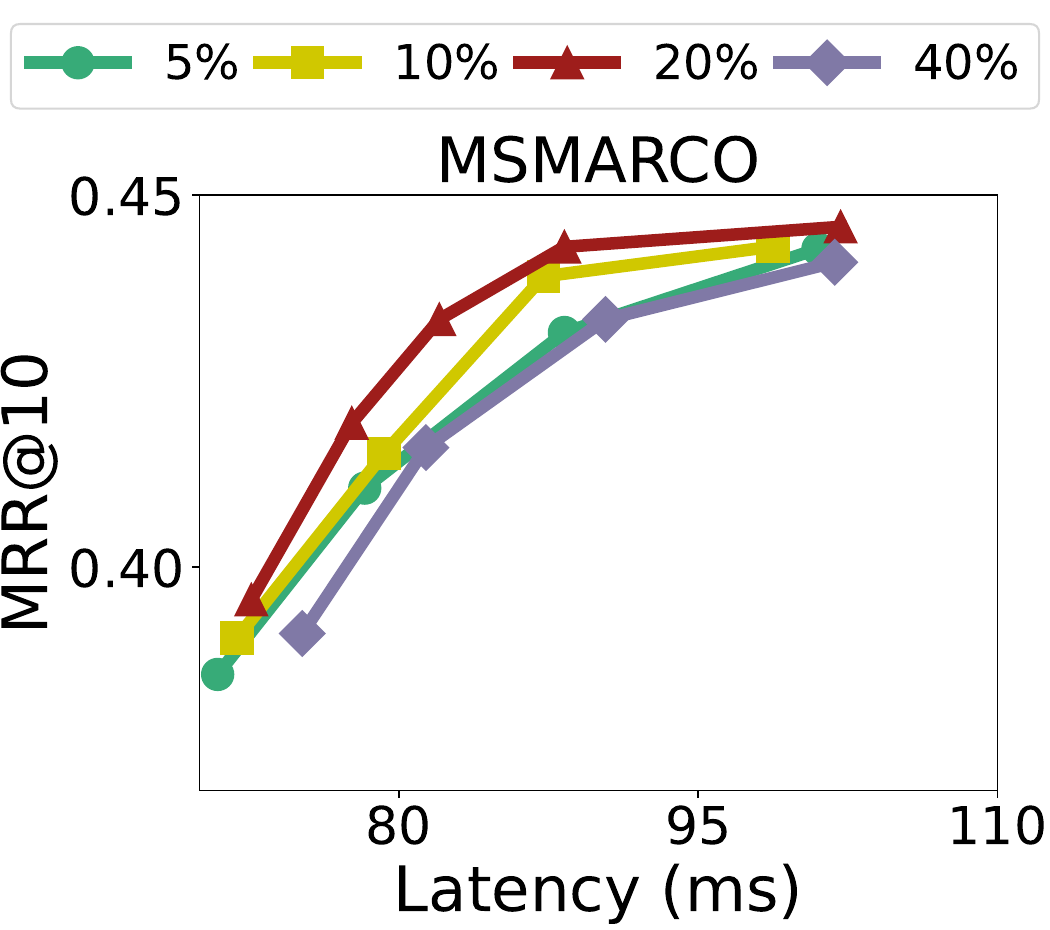}
    \caption{\rone{Impact of Shortcuts}}
    \vspace{-2ex}
    \label{fig:shortcuts}
  \end{minipage}
  \begin{minipage}[t]{0.235\textwidth}
    \centering
    \includegraphics[width=\textwidth]{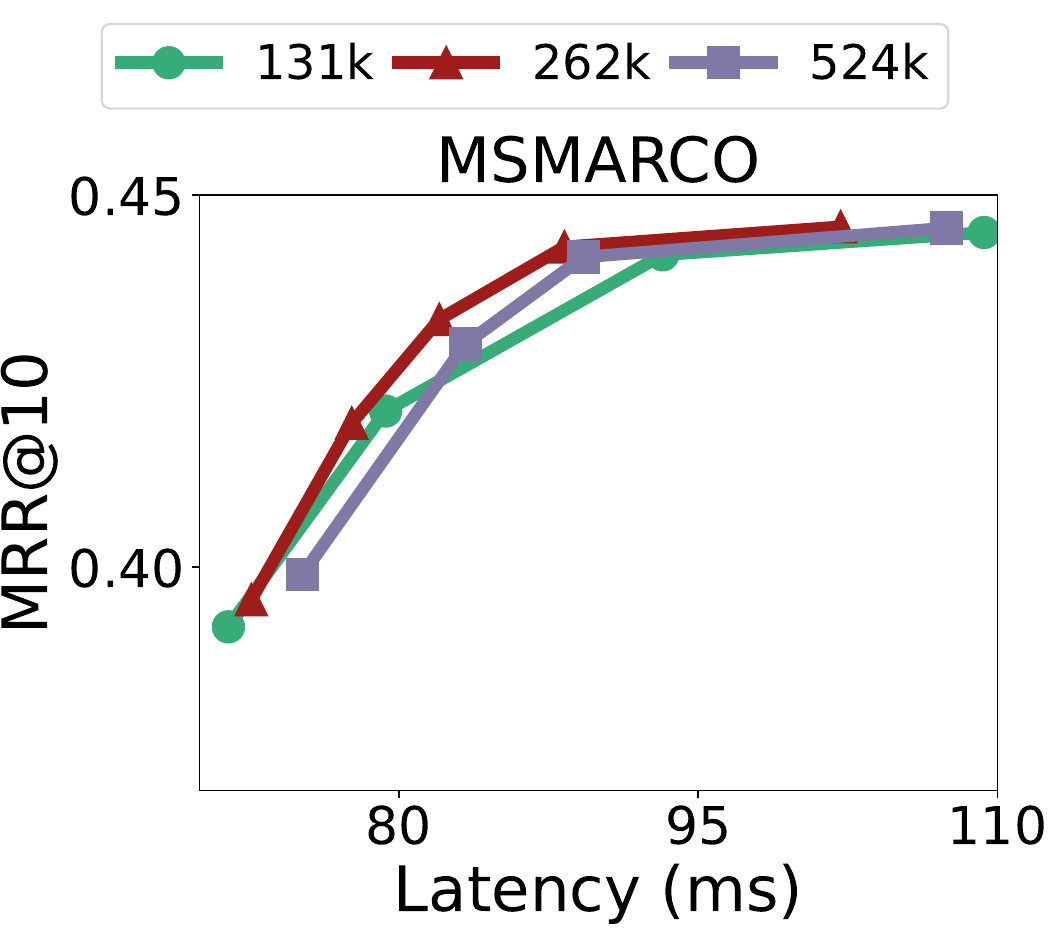}
    \caption{\rone{Impact of $|\mathcal{C}_\text{quant}|$}}
    \label{fig:cquant}
    \vspace{-2ex}
  \end{minipage}
\end{figure}

\subsubsection{\textbf{\rone{Effect of Shortcut Coverage.}}}
\rone{In this experiment, we study how the amount of training data used for shortcut construction affects query performance. We randomly sample 5\%, 10\%, 20\%, and 40\% of the training instances, resulting in approximately 10k, 20k, 40k, and 80k shortcut edges, respectively.
As shown in Figure~\ref{fig:shortcuts}, increasing the sampling ratio from 5\% to 20\% yields a better MRR–latency trade-off, whereas further increasing it to 40\% degrades performance. 
The recall–latency and success–latency curves are omitted as they follow almost the same trend.
Such results indicate that sparse shortcuts provide insufficient long-range connectivity, limiting search performance for some hard-to-find cases. 
In contrast, excessive shortcuts densify the graph, 
forcing the search to perform more distance computations per step during traversal, thereby offsetting the gains from improved connectivity. 
Therefore, we set the sampling ratio to 20\% by default, which provides a favorable balance between accuracy and efficiency.}

\subsubsection{\textbf{\rone{Effect of $|\mathcal{C}_\text{quant}|$.}}}
\rone{
In this experiment, we evaluate the effect of the number of quantization centroids $|C_{\text{quant}}|$ on retrieval performance by varying $|C_{\text{quant}}| \in \{131\text{k}, 262\text{k}, 524\text{k}\}$. Recall that $262k$ is the default value for MSMARCO derived from the empirical formula in Section \ref{sec:para}.
As shown in Figure~\ref{fig:cquant}, the retrieval performance of GEM is robust to the choices of $|C_{\text{quant}}|$.
Although increasing $|C_{\text{quant}}|$ leads to higher distance computation overhead, it yields more accurate approximations and thus requires fewer candidates for reranking. In contrast, reducing $|C_{\text{quant}}|$ lowers the computational cost but necessitates probing more candidates to achieve a comparable  accuracy level. Therefore, we simply adopt the empirical formula suggested by prior work to set the default value of $|C_{\text{quant}}|$. 
}

%% file: sec-conclusion.tex
\section{Conclusion}\label{sec:conclusion}



We introduced GEM, a native graph-based framework designed for multi-vector retrieval. 
GEM constructs a set-level dual-graph with metric decoupling to represent complex inter-set relationships, and further refines it through adaptive TF-IDF–guided clustering and semantic shortcuts.
By incorporating quantized distance estimation and cluster-guided multi-path search, GEM jointly enhances efficiency and accuracy, effectively bridging the gap between semantic expressiveness and retrieval efficiency.
Extensive experiments showed that GEM consistently surpasses state-of-the-art methods in retrieval quality and latency, while maintaining comparable or smaller index size and construction cost. These results confirmed GEM as a practical solution for large-scale multi-vector retrieval.  For future work, integrating ultra-low-bit quantization could further reduce computational and memory costs, paving the way for broader industrial adoption.